\documentclass[pra,aps,superscriptaddress,showpacs,onecolumn]{revtex4}
\usepackage{amssymb}
\usepackage{amsmath}
\usepackage{amssymb}
\usepackage{graphicx}
\usepackage{bbm}
\usepackage{afterpage}%
\usepackage{epsfig}
\usepackage{amsfonts}
\def\QED{\leavevmode\unskip\penalty9999 \hbox{}\nobreak\hfill
     \quad\hbox{\leavevmode  \hbox to.77778em{%
               \hfil\vrule   \vbox to.675em%
               {\hrule width.6em\vfil\hrule}\vrule\hfil}}
     \par\vskip3pt}
\def\qed{\leavevmode\unskip\penalty9999 \hbox{}\nobreak\hfill
     \quad\hbox{\leavevmode  \hbox to.77778em{%
               \hfil\vrule   \vbox to.675em%
               {\hrule width.6em\vfil\hrule}\vrule\hfil}}
\par\vskip3pt}
\def\ibb #1{\leavevmode\hbox{\kern.3em\vrule
     height 1.5ex depth -.1ex width .4pt\kern-.3em\rm#1}}

\newcommand{\be}{\begin{equation}}
\newcommand{\ee}{\end{equation}}
\newcommand{\ba}{\begin{array}}
\newcommand{\ea}{\end{array}}
\newcommand{\bqa}{\begin{eqnarray}}
\newcommand{\eqa}{\end{eqnarray}}

\providecommand{\U}[1]{\protect\rule{.1in}{.1in}}
\providecommand{\U}[1]{\protect\rule{.1in}{.1in}}
\providecommand{\U}[1]{\protect\rule{.1in}{.1in}}
\newtheorem{theorem}{Theorem}[section]

\newtheorem{example}[theorem]{Example}

\begin{document}

\title{Improved lower bounds on genuine-multipartite-entanglement concurrence}
\author{Zhi-Hua Chen}
\affiliation {Department of Science, Zhijiang college, Zhejiang
University of technology, Hangzhou, 310024, P.R.China.}
\author{ Zhi-Hao Ma}
\email{ma9452316@gmail.com}
\affiliation {Department of Mathematics, Shanghai Jiao-Tong University,
Shanghai, 200240, P. R. China; currently visiting the Department of Physics \&
Astronomy, University College London, WC1E 6BT London, United Kingdom;
}
\author{ Jing-Ling Chen}
\affiliation{Theoretical Physics Division, Chern Institute of
Mathematics, Nankai University, Tianjin, 300071, P. R.China; Centre for
Quantum Technologies, National University of Singapore, 3 Science Drive 2,
117543 }
\author{Simone Severini}
\affiliation{Department of Computer Science, and Department of
Physics \& Astronomy, University College London, WC1E 6BT London, United
Kingdom.}

\begin{abstract}
Genuine-multipartite-entanglement (GME) concurrence is a measure of genuine
multipartite entanglement that generalizes the well-known notion of
concurrence. We define an observable for GME concurrence. The observable
permits us to avoid full state tomography and leads to different analytic
lower bounds. By means of explicit examples we show that entanglement criteria
based on the bounds have a better performance with respect to the known methods.

\end{abstract}

\maketitle

\section{Introduction}

Entanglement plays a fundamental role in the study of many-body quantum
mechanics. Complex systems with multipartite quantum correlations have been
shown to be useful in numerous tasks, ranging from measurement based quantum
computing \cite{Raussendorf}, quantum secret sharing \cite{Schauer}, quantum
communication \cite{Markham}, \emph{etc.} Compared with the bipartite case,
multipartite entanglement is well-known to exhibit richer structures and a
variety of classes (see, \emph{e.g.}, \cite{Dur00, Barreiro10}). Being
substantially different from partial entanglement (\emph{i.e.}, entanglement
specified by correlations between any two subsystems), the so-called
\emph{genuine multipartite entanglement} (for short, \emph{GME}) is of special
interest. Although many efforts have been devoted towards the detection of GME
(\emph{e.g.}, entanglement witnesses \cite{Guhe11}, Bell-like inequalities
\cite{Bell}, \emph{etc.}), its characterization still remains a difficult
problem \cite{Horodecki09}. On the other hand, the quantitative aspects are
important because these are justified by the experimental perspective
\cite{Guhne09}.

While the three-tangle is a famous measure of GME for three qubits
\cite{Coffman00}, no such a concept is currently available for systems of
higher dimension. Recently, a notion of generalized concurrence, called
\emph{GME-concurrence }\cite{Ma11}, was introduced in the attempt of
distinguishing between GME and partial entanglement. In the present paper, we
first point out that GME concurrence of pure states may be directly accessible
in laboratory experiments (\emph{i.e.}, no full state tomography is needed)
provided that a two-fold copy of the state is available. Then, we present some
explicit lower bounds. We illustrate detailed examples in which the given
bounds perform better when compared with other known detection criteria. The
evaluation of the bounds permit to bypass full quantum tomography, since we
only need a polynomial number (on the system's dimension) of expectation
values. The results appear to be an improvement over\emph{ }\cite{Ma11}.

The remainder of the paper is organized as follows. In Section 2, we recall
the definition and the basic properties of GME-concurrence. In Section 3, we
show that GME-concurrence is an observable measure. In Section 4, we state and
prove the bounds. Section 5 is devoted to examples. Section 6 contains some
brief conclusions.

\section{GME-concurrence}

An $N$-partite pure state $|\phi\rangle$ with Hilbert space $\mathcal{H}%
_{1}\otimes\mathcal{H}_{2}\otimes\cdots\otimes\mathcal{H}_{N}$ is said to be
\emph{biseparable} if there is a bipartition $\gamma|\gamma^{^{\prime}}$ such
that $|\phi\rangle$ can be decomposed as a tensor product $|\phi
_{\gamma|\gamma^{^{\prime}}}\rangle=|\phi_{\gamma}\rangle\otimes|\phi
_{\gamma^{^{\prime}}}\rangle$. If an $N$-partite pure state is not biseparable
then it is said to be \emph{genuinely }$N$\emph{-partite entangled}. The same
terms apply to an $N$-partite mixed state $\rho$, if it can (resp. it can
not)\ be written as a convex combination of biseparable pure states $\rho
=\sum\limits_{i}p_{i}|\phi_{\gamma_{i}|\gamma_{i}^{^{\prime}}}\rangle
\langle\phi_{\gamma_{i}|\gamma_{i}^{^{\prime}}}|$, where each component
$|\phi_{\gamma_{i}|\gamma_{i}^{^{\prime}}}\rangle$ is biseparable (possibly
under different partitions). Given an $N$-partite pure state $|\phi\rangle$,
let $\gamma=\{j_{1},j_{2},...,j_{k}\}\subseteq\{1,2,...,N\}$ be a subset
inducing a bipartition $j_{1},j_{2},\ldots,j_{k}|j_{k+1},\ldots,j_{N}$. If
$C_{\gamma}^{2}(\phi):=1-Tr(\rho_{\gamma}^{2})$, where $\rho_{\gamma}$ is the
reduced density matrix of the subsystem indexed by $\gamma$, the
GME-concurrence (of a pure state) is%
\[
C_{GME}(\phi):=\sqrt{\min\limits_{\gamma}C_{\gamma}^{2}(\phi)}.
\]

For example, let us consider a three-qubit state $|\phi\rangle$. In this case,
we have $\gamma=\{1\}$, $\gamma=\{2\}$ or $\gamma=\{3\}$, corresponding to the
partitions $1|2,3$, $2|1,3$, and $3|1,2$, respectively. Its GME-concurrence is
then%
\[
C_{GME}^{2}(\phi)=\min\limits_{\gamma=\{1\},\{2\},\{3\}}\{1-Tr(\rho_{1}%
^{2}),1-Tr(\rho_{2}^{2}),1-Tr(\rho_{3}^{2})\}.
\]

More generally, the \emph{GME-concurrence} of an $N$-partite mixed state
$\rho$ is%
\begin{equation}
C_{GME}(\rho):=\min\sum_{i}p_{i}C_{GME}(\phi_{i}), \label{convex}%
\end{equation}
where the minimum is taken over all pure states decompositions $\rho=\sum
_{i}p_{i}|\phi_{i}\rangle\langle\phi_{i}|$. It is worth recalling that
GME-concurrence satisfies the following useful properties \cite{Ma11}:

\begin{description}
\item[M1.] The GME-concurrence is zero for all biseparable states;

\item[M2.] The GME-concurrence is strictly greater than zero for all GME states;

\item[M3.] (Convexity) $C_{GME}(\sum_{i}p_{i}\rho_{i})\leq\sum_{i}p_{i}%
C_{GME}(\rho_{i})$;

\item[M4.] (Non-increasing under LOCC) $C_{GME}(\Lambda_{LOCC}(\rho))\leq
C_{GME}(\rho)$;

\item[M5.] (Invariance under local unitary transformations) $C_{GME}%
(U_{local}\rho U_{local}^{+})=C_{GME}(\rho)$;

\item[M6.] (Subadditivity)$\ C_{GME}(\rho\otimes\sigma)\leq C_{GME}%
(\rho)+C_{GME}(\sigma)$.
\end{description}

\section{GME-concurrence of pure states is observable}

In this section, we describe an observable for GME-concurrence of pure states.
More specifically, GME-concurrence for pure state can be measured directly,
provided that two copies of the state are available. Notice that our approach
is quite different from that of \cite{Aolita06}, in particular, we do not use
symmetric and antisymmetric projections subspace. As above, let $|\phi\rangle$
be an $N$-partite pure state with Hilbert space $\mathcal{H}_{1}%
\otimes\mathcal{H}_{2}\otimes\cdots\otimes\mathcal{H}_{N}$, of respective
dimensions $d_{1},d_{2},...,d_{N}$. We can write
\begin{equation}
|\phi\rangle:=\sum\limits_{i_{1},i_{2},...,i_{N}}\phi_{i_{1}i_{2}\cdots i_{N}%
}|i_{1}i_{2}\cdots i_{N}\rangle, \label{pure}%
\end{equation}
where $i_{j}$ is the $i$-th element of an orthonormal basis of $\mathcal{H}%
_{j}$, with $j=1,...,N$. Given a subset $\gamma=\{t_{1},t_{2},...,t_{k}%
\}\subseteq\{1,2,...,N\}$, the $\gamma$-concurrence of $\rho$ can be written
as $C_{\gamma}^{2}(\phi)=\langle\phi|\otimes\langle\phi|B_{\gamma}|\phi
\rangle\otimes|\phi\rangle$. The observable $B_{\gamma}$ is independent of
$|\phi\rangle\ $and it is uniquely determined by the partition induced by
$\gamma$. The definition of $B_{\gamma}$ requires some preparation. Let
$\gamma=\{t_{1},t_{2},...,t_{k}\}$. Let%
\[%
\begin{tabular}
[c]{lll}%
$I:=\{j_{1},...,j_{k}\}\cup\{j_{k+1},j_{k+2},...,j_{N}\}$ & and &
 $J^{\prime
}:=\{j_{1}^{\prime},...,j_{k}^{\prime}\}\cup\{j_{k+1}^{\prime},j_{k+2}%
^{\prime},...,j_{N}^{\prime}\}$%
\end{tabular}
\
\]
be two arbitrary index sets such that $j_{i},j_{i}^{\prime}=0,...,d_{t_{i}}%
-1$, with $t_{i}=1,...,N$. Here, $\{j_{1},...,j_{k}\}$ and $\{j_{1}^{\prime
},...,j_{k}^{\prime}\}$ indicate the same positions as the ones indexed by
$\gamma$. For instance, for a three-qubit state, if $\gamma=\{2\}$ then
$\{j_{1},...,j_{k}\}$ and $\{j_{1}^{\prime},...,j_{k}^{\prime}\}$ indicate
elements in the second subsystem $\mathcal{H}_{2}$. We then define two further
index subsets,
\[%
\begin{tabular}
[c]{lll}%
$I_{\gamma}:=\{j_{1},...,j_{k}\}$ & and & $J_{\gamma}^{\prime}:=\{j_{1}%
^{\prime},...,j_{k}^{\prime}\}$.
\end{tabular}
\
\]
These subsets are obtained from $I$, $J^{\prime}$ and the partition $\gamma$.
We also define the complements
\[%
\begin{tabular}
[c]{lll}%
$I\backslash I_{\gamma}:=\{j_{k+1},j_{k+2},...,j_{N}\}$ & and & $J^{\prime
}\backslash J_{\gamma}^{\prime}:=\{j_{k+1}^{\prime},j_{k+2}^{\prime}%
,...,j_{N}^{\prime}\}$.
\end{tabular}
\
\]
Finally, we have the following two index sets obtained by swapping the
elements in the positions corresponding to the ones indexed by $\gamma$:
$I^{\prime}:=J_{\gamma}^{\prime}\cup I\backslash I_{\gamma}$ and

$J:=I_{\gamma}\cup J^{\prime}\backslash J_{\gamma}^{\prime}$.

Once $I$ and
$J^{\prime}$ are arbitrarily fixed, then $I^{\prime}$ and $J$ are uniquely
determined by $\gamma$. With the use of this notation, we can finally write
\begin{align*}
C_{\gamma}^{2}(\phi)  &  =1-Tr(\rho_{\gamma}^{2})  =\left(  \sum\limits_{i_{1},i_{2},...,i_{N}}\phi_{i_{1}i_{2}\cdots i_{N}%
}\bar{\phi}_{i_{1}i_{2}\cdots i_{N}}\right)  ^{2}\\
&-\sum\limits_{I_{\gamma}}%
\sum\limits_{J_{\gamma}^{\prime}}\sum\limits_{I\backslash I_{\gamma}}\phi
_{I}\bar{\phi}_{I^{\prime}}\sum\limits_{J^{\prime}\backslash J_{\gamma
}^{\prime}}\phi_{J^{\prime}}\bar{\phi}_{J}\\
&  =\sum\limits_{I,J^{\prime}}\phi_{I}\bar{\phi}_{I}\phi_{J^{\prime}}\bar
{\phi}_{J^{\prime}}-\sum\limits_{I,J^{\prime}}\phi_{I}\bar{\phi}_{I^{\prime}%
}\phi_{J^{\prime}}\bar{\phi}_{J}\\
&  =\sum\limits_{I,J^{\prime}}(\phi_{I}\phi_{J^{\prime}}-\phi_{I^{\prime}}%
\phi_{J})(\bar{\phi}_{I}\bar{\phi}_{J^{\prime}}-\bar{\phi}_{I^{\prime}}%
\bar{\phi}_{J})\\
&  =\sum\limits_{I,J^{\prime}}|(\phi_{I}\phi_{J^{\prime}}-\phi_{I^{\prime}%
}\phi_{J})|^{2},
\end{align*}
where the sum is taken over all possible index sets $I$ and $J^{\prime}$. The
observable is
\begin{align*}
B_{\gamma}  &  =\sum\limits_{i_{1},i_{2},...,i_{N}}|i_{1}i_{2}\cdots
i_{N}\rangle\otimes|i_{1}i_{2}\cdots i_{N}\rangle\\
&\times\sum\limits_{i_{1}%
,i_{2},...,i_{N}}\langle i_{1}i_{2}\cdots i_{N}|\otimes\langle i_{1}%
i_{2}\cdots i_{N}|\\
&  -\sum\limits_{I_{\gamma}}\sum\limits_{J_{\gamma}^{\prime}}\sum
\limits_{I\backslash I_{\gamma}}|I\rangle\otimes|I^{\prime}\rangle
\sum\limits_{I\backslash I_{\gamma}}\langle I|\otimes\langle I^{\prime}|.
\end{align*}
It follows that
\[
C_{\gamma}^{2}(\phi)=\sum\limits_{I,J^{\prime}}|(\phi_{I}\phi_{J^{\prime}%
}-\phi_{I^{\prime}}\phi_{J})|^{2}=\langle\phi|\otimes\langle\phi|B_{\gamma
}|\phi\rangle\otimes|\phi\rangle,
\]

which gives a general expression for the GME-concurrence of a pure state:

\begin{align*}
C_{GME}^{2}(\phi)&=\min\limits_{\gamma}\langle\phi|\otimes\langle\phi
|B_{\gamma}|\phi\rangle\otimes|\phi\rangle\\
&=\min\limits_{\gamma}\left(
\sum\limits_{I,J,I^{\prime},J^{\prime}}|(\phi_{I}\phi_{J^{\prime}}%
-\phi_{I^{\prime}}\phi_{J})|^{2}\right)  .
\end{align*}

For example, let\emph{ }%
\begin{equation}
|\phi\rangle=\sum_{i,j,k\in\{0,1\}}\phi_{ijk}|ijk\rangle\label{ps}%
\end{equation}
be a generic three-qubit pure state. If $\gamma=\{1\}$ then
\begin{align*}
C_{1}^{2}(\phi)  &  =\langle\phi|\otimes\langle\phi|B_{1}|\phi\rangle
\otimes|\phi\rangle\\
&  =2(|\phi_{000}\phi_{101}-\phi_{100}\phi_{001}|^{2}+|\phi_{000}\phi
_{110}-\phi_{100}\phi_{010}|^{2}\\&+|\phi_{000}\phi_{111}-\phi_{100}\phi
_{011}|^{2}\\
&  +|\phi_{001}\phi_{110}-\phi_{101}\phi_{010}|^{2}+|\phi_{001}\phi
_{111}-{\phi}_{101}{\phi}_{011}|^{2}\\&+|\phi_{010}\phi_{111}-{\phi}_{110}{\phi
}_{011}|^{2}),
\end{align*}
and the observable is%
\begin{align*}
B_{1}&=\sum\limits_{i,j,k\in\{0,1\}}|ijk|ijk\rangle\sum\limits_{i,j,k\in
\{0,1\}}\langle ijk|ijk|\\&-\sum\limits_{i\in\{0,1\}}\sum\limits_{i^{\prime}%
\in\{0,1\}}\sum\limits_{j,k}|ijk|i^{\prime}jk\rangle\sum\limits_{j,k}\langle
ijk|i^{\prime}jk|.
\end{align*}
The observables $B_{2}$ and $B_{3}$ are obtained analogously.

\section{Lower bounds}

\subsection{Statement of the results\label{res}}

Let $|\psi\rangle=\bigotimes_{i=1}^{N}|x_{i}\rangle=|x_{1}x_{2}\cdots
x_{N}\rangle$ be a product state with Hilbert space $\mathcal{H}%
=\mathcal{H}_{1}\otimes\mathcal{H}_{2}\otimes\cdots\otimes\mathcal{H}_{N}$.
Let $|\psi_{i}\rangle=|x_{1}x_{2}\cdots x_{i-1}x_{i}^{\prime}x_{i+1}\cdots
x_{N}\rangle$ and $|\psi_{j}\rangle=|x_{1}x_{2}\cdots x_{j-1}x_{j}^{\prime
}x_{j+1}\cdots x_{N}\rangle$ be the product states obtained from $|\psi
\rangle$ by applying (independently) local unitaries to $|x_{i}\rangle
\in\mathcal{H}_{i}$ and $|x_{j}\rangle\in\mathcal{H}_{j}$, respectively. Let
$|\Psi_{ij}\rangle:=|\psi_{i}\rangle|\psi_{j}\rangle$ be a product state on
$\mathcal{H}^{\otimes2}=(\mathcal{H}_{1}\otimes\mathcal{H}_{2}\otimes
\cdots\otimes\mathcal{H}_{N})^{\otimes2}$. Let us define $\Pi=\mathcal{P}%
_{1}\circ\mathcal{P}_{2}\circ\cdots\circ\mathcal{P}_{N}$, where $\mathcal{P}%
_{i}$ is the operator swapping the two copies of $\mathcal{H}_{i}$ in
$\mathcal{H}^{\otimes2}$, for each $i=1,...,N$. Finally, let $\rho$ be an
arbitrary state in the total Hilbert space $\mathcal{H}$. We will prove the
following statements:

$\bf{Bound} 1.$ By writing%

\begin{align}
\mathcal{F}(\rho,\psi)  &  =\sum\limits_{1\leq i\neq j\leq N}\sqrt{\langle
\Psi_{ij}|\rho^{\otimes2}\Pi|\Psi_{ij}\rangle}\nonumber
\\&-\sum\limits_{1\leq i\neq j\leq
N}\sqrt{\langle\Psi_{ij}|\mathcal{P}_{i}^{\dagger}\rho^{\otimes2}%
\mathcal{P}_{i}|\Psi_{ij}\rangle}\label{bound1}\\
&  -(N-2)\sum\limits_{1\leq i\leq N}\sqrt{\langle\Psi_{ii}|\mathcal{P}%
_{i}^{\dagger}\rho^{\otimes2}\mathcal{P}_{i}|\Psi_{ii}\rangle},\nonumber
\end{align}

we have%
\begin{equation}
\mathcal{F}(\rho,\psi)\leq\sqrt{2}{(N-1)}\cdot C_{GME}(\rho). \label{result}%
\end{equation}

$\bf{Bound} 2.$ For any given $|\psi\rangle$,we can get $\psi_{i}=|x_{1}\cdots
x_{i-1}x_{i}^{\prime}x_{i+1}\cdots x_{j-1}x_{j}x_{j+1}\cdots x_{N}\rangle$ by
changing the $i$ th bit of $|\psi\rangle$,Let $|\psi_{ij}\rangle=|x_{1}\cdots
x_{i-1}x_{i}^{\prime}x_{i+1}\cdots x_{j-1}x_{j}^{\prime}x_{j+1}\cdots
x_{N}\rangle$ by changing the $j$th bit of $\psi_{i}\rangle$, and $\Psi
_{i_{l}i_{m}}:=|\psi_{i_{l}}\rangle|\psi_{i_{m}}\rangle$ be defined as above,
but obtained by the application of two local unitaries.
By writing
\begin{align*}
\mathcal{L}(\rho,\psi_{i}) & = \sum\limits_{l\neq m,l\neq i,m\neq i}%
\sqrt{\langle\Psi_{i_{l}i_{m}}|\rho^{\otimes2}\Pi|\Psi_{i_{l}i_{m}}\rangle
}\\ & \nonumber-\sum\limits_{l\neq m,l\neq i,m\neq i}\sqrt{\langle\Psi_{i_{l}i_{m}%
}|\mathcal{P}_{l}^{\dagger}\rho^{\otimes2}\mathcal{P}_{l}|\Psi_{i_{l}i_{m}%
}\rangle}\\
&  -(N-3)\sum\limits_{l\neq i}\sqrt{\langle\Psi_{i_{l}i_{l}}|\mathcal{P}%
_{l}^{\dagger}\rho^{\otimes2}\mathcal{P}_{l}|\Psi_{i_{l}i_{l}}\rangle},
\end{align*}
we have%
\begin{equation}
\sum\limits_{1\leq i\leq N}\mathcal{L}(\rho,\psi_{i})\leq2\sqrt{N-2}\cdot
C_{GME}(\rho). \label{eqth1}%
\end{equation}

$\bf{Bound} 3.$ Let $V=\{|\chi_{1}\rangle,...,|\chi_{m}\rangle\}$ be a set of
product states in $\mathcal{H}$. Then%
\begin{align}
\mathcal{T}(\rho,\chi)  & \nonumber  =\sum\limits_{|\chi_{\alpha\rangle\in V}}\sum\limits_{|\chi_{\beta}\rangle\in K_{\alpha}}(|\langle\chi_{\alpha}%
|\rho|\chi_{\beta}\rangle|\\& \nonumber -\sqrt{\langle\chi_{\alpha}|\otimes\langle
\chi_{\beta}|\Pi_{\alpha\beta}\rho^{\otimes2}\Pi_{\alpha\beta}|\chi_{\alpha
}\rangle\otimes|\chi_{\beta}\rangle})\\& \nonumber
-(s-s_{0})\sum\limits_{\alpha}\langle\chi_{\alpha}|\rho|\chi_{\alpha}\rangle
\\&  \leq\sqrt{2}s\cdot C_{GME}(\rho), \label{eqth2}%
\end{align}
where%
\[
K_{\alpha}=\{|\chi_{\beta}\rangle:||\chi_{\alpha}\rangle\cap|\chi_{\beta
}\rangle|=N-2\text{ with }|\chi_{\alpha}\rangle,|\chi_{\beta}\rangle\in V\},
\]
and $s=\max{|K_{\alpha}|}$. Additionally,
\[
s_{0}=\min\limits_{1\leq i\leq N}s_{\alpha,i},
\]
where $s_{\alpha,i}$ is the number of the vectors in $K_{\alpha}$ such that
the $i$-th bits of $v_{\alpha}$ are different when $K_{\alpha}\neq\emptyset$.
We denote by $||v_{\alpha}\rangle\cap|v_{\beta}\rangle|$ the number of
coordinates that are equal in both vectors.$\Pi_{\alpha\beta}$ swaps one
different bit of $|\chi_{\alpha}\rangle$ and $|\chi_{\beta}\rangle$.

The definition of the witness in Bound 1 appeared in \cite{Huber10,Guhne}. When
$m=2$ and $N=4$ the bound in Eq. (\ref{eqth1}) is the same as the criterion
given in \cite{Huber11a}. However, this is not always the case, as we shall
verify below.

\subsection{Proof}

\subsubsection{Bound 1}

We start with a three-qubit state to get an intuition for the general case
that we shall discuss later. We are interested in bounding the GME concurrence
of an arbitrary three-qubit state $\rho=\sum_{i}p_{i}\phi^{(i)}$, with pure
state decomposition $\{p_{i},\phi^{(i)}\}$. We select a product state
$|\psi\rangle=|001\rangle$. If we apply the bit flip operation to the $i$-th
qubit, we have $|\psi_{1}\rangle=|101\rangle$, $|\psi_{2}\rangle=|011\rangle$,
and $|\psi_{3}\rangle=|000\rangle$. The bound given in Eq. (\ref{result}) is%
\begin{align}
\mathcal{F}(\rho,\psi)& \nonumber=2(|\rho_{4,6}|+|\rho_{1,4}|+|\rho_{1,6}|-\sqrt
{\rho_{2,2}\rho_{8,8}}\\&\nonumber-\sqrt{\rho_{2,2}\rho_{3,3}}\sqrt{\rho_{2,2}\rho_{5,5}%
})\\& -\rho_{1,1}-\rho_{4,4}-\rho_{6,6}
\leq2\sqrt{2}\cdot C_{GME}(\rho).
\label{fpsir}%
\end{align}
For proving this, let us consider the pure state $|\phi\rangle$ as in Eq.
(\ref{ps}). With the use of the Cauchy-Schwarz and the triangle inequality, we
obtain $C_{i}(\phi)$, for $i=1,2,3$:%

\begin{align*}
\sqrt{2}C_{1}(\phi)   \geq(|\phi_{011}\phi_{101}|-|\phi_{001}\phi
_{111}|+|\phi_{000}\phi_{101}|-|\phi_{001}\phi_{100}|)\\
\sqrt{2}C_{2}(\phi)  \geq(|\phi_{000}\phi_{011}|-|\phi_{001}\phi
_{010}|+|\phi_{011}\phi_{101}|-|\phi_{001}\phi_{111}|)\\
\sqrt{2}C_{3}(\phi)    \geq(|\phi_{000}\phi_{011}|-|\phi_{001}\phi
_{010}|+|\phi_{000}\phi_{101}|-|\phi_{001}\phi_{100}|)
\end{align*}

By the same step,
\begin{align*}
\mathcal{F}(\phi,\psi)&  =2(|\phi_{011}\phi_{101}|-|\phi_{001}\phi_{111}%
|+|\phi_{000}\phi_{011}|\\& \nonumber-|\phi_{001}\phi_{010}|+|\phi_{000}\phi_{101}%
|-|\phi_{001}\phi_{100}|)\\& \nonumber
-(|\phi_{000}|^{2}+|\phi_{011}|^{2}+|\phi_{101}|^{2})\\&
\leq2\sqrt{2}\min
\{C_{1}(\phi),C_{2}(\phi),C_{3}(\phi)\}.
\end{align*}
This confirms the statement in Eq. (\ref{fpsir}), when restricted to pure
states. If $\rho$ is a mixed state, the convex roof construction is bounded
as
\[
2\sqrt{2}C_{GME}(\rho)\geq\inf\limits_{\{p_{i},|\phi_{i}\rangle\}}\sum
_{i}p_{i}\mathcal{F}(\phi^{(i)},\psi),
\]
where ${\{}p_{i},\phi^{(i)}{\}}$ is any pure state decomposition of $\rho$.
Having chosen $|\psi\rangle=|001\rangle$, we obtain Eq. (\ref{fpsir}). Since
$C_{GME}(\rho)$ is invariant under local unitaries, for any choice of a
product state $|\psi\rangle$, $\mathcal{F}(\rho,\psi)$ is a lower bound to
$C_{GME}(\rho)$ leading to $C_{GME}(\rho)\geq\frac{1}{2}\mathcal{F}(\rho
,\psi)$. This concludes the proof for the three-qubit case.

We are now ready to prove the inequality for a general $N$-qudit state. Some
notation is needed:%
\begin{align*}
c_{0}  &  :=x_{1}\cdots x_{i-1}x_{i}x_{i+1}\cdots x_{N};\\
c_{i}  &  :=x_{1}\cdots x_{i-1}x_{i}^{\prime}x_{i+1}\cdots x_{N};\\
c_{j}  &  :=x_{1}\cdots x_{j-1}x_{j}^{\prime}x_{j+1}\cdots x_{N};\\
c_{ij}  &  :=x_{1}\cdots x_{i-1}x_{i}^{\prime}x_{i+1}\cdots x_{j-1}%
x_{j}^{\prime}x_{j+1}\cdots x_{N};
\end{align*}
when $i<j$, we use $c_{ij}$; otherwise, we use $c_{ji}$.

Again, since $C_{GME}(\rho)$ is invariant under local unitaries, we only need
to consider the integers $0\leq x_{i}\leq d_{i}-1$, for $i=1,2,...,N$. For the
generic $N$-qudit pure state $|\phi\rangle$ in Eq. (\ref{pure}), the bound in
Eq. (\ref{bound1}) reads as
\[
\mathcal{F}(\phi,\psi)=\sum\limits_{1\leq i\neq j\leq N}(|\phi_{c_{i}}%
\phi_{c_{j}}|-|\phi_{c_{0}}\phi_{c_{ij}}|)-(N-2)\sum\limits_{1\leq i\leq
N}|\phi_{c_{i}}|^{2}.
\]
There are two cases depending on the biseparable partition $\gamma$:

\noindent\textbf{Case 1.}\emph{ }For any given $\gamma\subset\{1,2,...N\}$
with $|\{\gamma\}|=1$,%
\begin{align*}
2\sqrt{(N-1)}C_{\gamma}(\phi)  &  =2\sqrt{(N-1)}\sqrt{\sum\limits_{j\neq
\gamma}|\phi_{c_{\gamma}}\phi_{c_{j}}-\phi_{c_{0}}\phi_{c_{\gamma j}}|^{2}}\\
&  \geq\sum\limits_{j\neq\gamma}|\phi_{c_{\gamma}}\phi_{c_{j}}-\phi_{c_{0}%
}\phi_{c_{\gamma j}}| \\&
\geq\sum\limits_{j\neq\gamma}|\phi_{c_{\gamma}}%
\phi_{c_{j}}|-|\phi_{c_{0}}\phi_{c_{\gamma j}}|,
\end{align*}
It is convenient to interpret $\mathcal{F}(\phi,\psi)$ as a sum of two terms:%
\begin{align*}
\mathcal{F}(\phi,\psi)&=\sum\limits_{j\neq\gamma}(|\phi_{c_{j}}\phi_{c_{\gamma
}}|-|\phi_{c_{0}}\phi_{c_{j\gamma}}|)\\& +\sum\limits_{i\neq\gamma,j\neq\gamma
}(|\phi_{c_{i}}\phi_{c_{j}}|-|\phi_{c_{0}}\phi_{c_{ij}}|)-(N-2)\sum
\limits_{1\leq i\leq N}|\phi_{c_{i}}|^{2}\\&=X+Y,
\end{align*}
where%
\begin{align*}
X&=\sum\limits_{j\neq\gamma}(|\phi_{c_{j}}\phi_{c_{\gamma}}|-|\phi_{c_{0}}%
\phi_{c_{j\gamma}}|)
\leq\sum\limits_{j\neq\gamma}|\phi_{c_{\gamma}}\phi
_{c_{j}}-\phi_{c_{0}}\phi_{c_{j\gamma}}|\\&
\leq2\sqrt{(N-1)}C_{\gamma}(\phi)
\end{align*}
and
\begin{align*}
Y  &  =\sum\limits_{i\neq\gamma,j\neq\gamma}(|\phi_{c_{i}}\phi_{c_{j}}%
|-|\phi_{c_{0}}\phi_{c_{ij}}|)-(N-2)\sum\limits_{1\leq i\leq N}|\phi_{c_{i}%
}|^{2}\\
&  \leq\sum\limits_{i\neq\gamma,j\neq\gamma}\frac{|\phi_{c_{i}}|^{2}%
+|\phi_{c_{j}}|^{2}}{2}-(N-2)\sum\limits_{1\leq i\leq N}|\phi_{c_{i}}|^{2}\\
&  =(N-2)\sum\limits_{1\leq i\leq N}|\phi_{c_{i}}|^{2}-(N-2)\sum\limits_{1\leq
i\leq N}|\phi_{c_{i}}|^{2}\\
&  =0.
\end{align*}
Hence,%
\[
\mathcal{F}(\phi,\psi)\leq\min\limits_{\gamma=1,2,\cdots,N}2\sqrt{(N-1)}\cdot
C_{\gamma}(\phi).
\]

\noindent\textbf{Case 2.}\emph{ }For any given $\gamma=\{j_{1},j_{2}%
,...,j_{k}\}\subset\{1,2,...N\}$, with $k\geq2$,
\begin{align*}
&2\sqrt{(N-1)}\cdot C_{\gamma}(\phi)  \\&  \geq2\sqrt{(N-k)k}\cdot C_{\gamma
}(\phi)\\
&  =2\sqrt{(N-k)k}\sqrt{\sum\limits_{l=1}^{k}\sum\limits_{j\neq j_{t}:1\leq
t\leq k} |\phi_{c_{j_{l}}}\phi_{c_{j}}-\phi_{c_{0}}\phi_{c_{j_{l}j}}|^{2}}\\
&  \geq\sum\limits_{l=1}^{k}\sum\limits_{j\neq j_{t}:1\leq t\leq k}%
|\phi_{c_{j_{l}}}\phi_{c_{j}}|-|\phi_{c_{0}}\phi_{c_{j_{l}j}}|.
\end{align*}

\begin{widetext}
As in the previous case,

\begin{align*}
\mathcal{F}(\phi,\psi)  &  =\sum\limits_{l=1}^{k}\sum\limits_{j\neq
j_{t}:1\leq t\leq k}(|\phi_{c_{j}}\phi_{c_{j_{l}}}|-|\phi_{c_{0}}%
\phi_{c_{jj_{l}}}|)\\&+\sum\limits_{i\neq j,i\neq j_{t},j\neq j_{t}:1\leq t\leq
k}(|\phi_{c_{i}}\phi_{c_{j}}|-|\phi_{c_{0}}\phi_{c_{ij}}|)\\
&  +\sum\limits_{l\neq t}(|\phi_{c_{j_{t}}}\phi_{c_{j_{l}}}|-|\phi_{c_{0}}%
\phi_{c_{j_{t}j_{l}}}|)-(N-2)\sum\limits_{i}|\phi_{c_{i}}|^{2}\\
&  =X+Y,
\end{align*}

where $X$ is the summand with $i=j_{l}$ and $j\neq j_{t}$ or $j=j_{l}$ and
$i\neq j_{t}$ ($1\leq t\leq k$ and $l=1,2,...,k$); $Y$ is the summand with
$i\neq j_{l}$ and $j\neq j_{l}$ or $i=j_{l},j=j_{t}$. Then,%
\[
X=\sum_{l=1}^{k}\sum\limits_{j\neq j_{t}:1\leq t\leq k}(|\phi_{c_{j}}%
\phi_{c_{j_{l}}}|-|\phi_{c_{0}}\phi_{c_{jj_{l}}}|)<2\sqrt{(N-1)}\cdot
C_{\gamma}(\phi)
\]
and%
\begin{align*}
Y  &  =\sum\limits_{i\neq j_{t},j\neq j_{t}:1\leq t\leq k}(|\phi_{c_{i}}%
\phi_{c_{j}}|-|\phi_{c_{0}}\phi_{c_{ij}}|)+\sum\limits_{l\neq t}%
(|\phi_{c_{j_{t}}}\phi_{c_{j_{l}}}|-|\phi_{c_{0}}\phi_{c_{j_{t}j_{l}}%
}|)-(N-2)\sum\limits_{i}|\phi_{c_{i}}|^{2}\\
&  \leq\sum\limits_{i\neq j_{t},j\neq j_{t}:1\leq t\leq k}\frac{|\phi_{c_{i}%
}|^{2}+|\phi_{c_{j}}|^{2}}{2}+\sum\limits_{l\neq t}\frac{|\phi_{c_{j_{l}}%
}|^{2}+|\phi_{c_{j_{t}}}|^{2}}{2}-(N-2)\sum\limits_{i}|\phi_{c_{i}}|^{2}\\
&  \leq(N-k-1)\sum\limits_{i\neq j_{t}:1\leq t\leq k}(|\phi_{c_{i}}%
|^{2})+(k-1)\sum\limits_{l}(|\phi_{c_{j_{l}}}|^{2})-(N-2)\sum\limits_{i}%
|\phi_{c_{i}}|^{2}\\
&  \leq(N-2)\sum\limits_{i}|\phi_{c_{i}}|^{2}-(N-2)\sum\limits_{i}|\phi
_{c_{i}}|^{2}\\
&  =0.
\end{align*}

Combining together the two cases above, we conclude that%
\[
\mathcal{F}(\phi,\psi)\leq\min\limits_{\gamma}2\sqrt{(N-1)}\cdot C_{\gamma
}(\phi)=2\sqrt{(N-1)}\cdot C_{GME}(\phi).
\]
This ends the proof of the result stated in Eq. (\ref{result}). The bound for
mixed stated is given directly by the convexity of the GME concurrence
(property M3 in Section 1) as follows. The bounds in Eqs. (\ref{eqth1}) and
Eq. (\ref{eqth2}) can\ be easily obtained in analogous way, as it will be
detailed in the next subsections.

Let
\begin{align*}
c_{0}  &  :=x_{1}\cdots x_{i-1}x_{i}x_{i+1}\cdots x_{N};\\
c_{i}  &  :=x_{1}\cdots x_{i-1}x_{i}^{\prime}x_{i+1}\cdots x_{N};\\
c_{j}  &  :=x_{1}\cdots x_{j-1}x_{j}^{\prime}x_{j+1}\cdots x_{N};\\
c_{ij}  &  :=x_{1}\cdots x_{i-1}x_{i}^{\prime}x_{i+1}\cdots x_{j-1}%
x_{j}^{\prime}x_{j+1}\cdots x_{N};
\end{align*}
when $i<j$, we use $c_{ij}$; otherwise, we use $c_{ji}$.

Let $\rho=\sum\limits_{i}t_{k}\rho^{(k)}$ be the optimal decomposition of
$\rho$ for the GME-concurrence, \emph{i.e}, $C_{GME}(\rho)=\sum\limits_{k}%
t_{k}\cdot C_{GME}(\rho^{(k)})$, then%

\begin{align*}
\mathcal{F}(\rho,\psi)  &  =\sum\limits_{1\leq i\neq j\leq N}|\rho
_{c_{i},c_{j}}|-\sum\limits_{1\leq i\neq j\leq N}\sqrt{\rho_{c_{0},c_{0}}%
\rho_{c_{ij},c_{ij}}}-(N-2)\sum\limits_{1\leq i\leq N}\rho_{c_{i},c_{i}}\\
&  =\sum\limits_{1\leq i\neq j\leq N}\left\vert \sum\limits_{k}t_{k}%
\rho_{c_{i},c_{j}}^{(k)}\right\vert -\sum\limits_{1\leq i\neq j\leq N}%
\sqrt{\sum\limits_{k}t_{k}\rho_{c_{0},c_{0}}^{(k)}\sum\limits_{k}t_{k}%
\rho_{c_{ij},c_{ij}}^{(k)}}-(N-2)\sum\limits_{1\leq i\leq N}\sum
\limits_{k}t_{k}\rho_{c_{i},c_{i}}^{(k)}%
\end{align*}
For the first term,%
\[
\sum\limits_{1\leq i\neq j\leq N}\left\vert \sum\limits_{k}t_{k}\rho
_{c_{i},c_{j}}^{(k)}\right\vert \leq\sum\limits_{k}t_{k}\sum\limits_{1\leq
i\neq j\leq N}|\rho_{c_{i},c_{j}}^{(k)}|.
\]
For the second term,
\[
\sqrt{\sum\limits_{k}t_{k}\rho_{c_{0},c_{0}}^{(k)}\sum\limits_{k}t_{k}%
\rho_{c_{ij},c_{ij}}^{(k)}}=\sqrt{\sum\limits_{k_{1},k_{2}}t_{k_{1}}t_{k_{2}%
}\rho_{c_{0},c_{0}}^{(k_{1})}\rho_{c_{ij},c_{ij}}^{(k_{2})}}\geq
\sum\limits_{k}t_{k}\sqrt{\rho_{c_{0},c_{0}}^{(k)}}\sqrt{\rho_{c_{ij},c_{ij}%
}^{(k)}}%
\]
Therefore
\[
\sum\limits_{1\leq i\neq j\leq N}\sqrt{\sum\limits_{k}t_{k}\rho_{c_{0},c_{0}%
}^{(k)}\sum\limits_{k}t_{k}\rho_{c_{ij},c_{ij}}^{(k)}}\geq\sum\limits_{1\leq
i\neq j\leq N}\sum\limits_{k}t_{k}\sqrt{\rho_{c_{0},c_{0}}^{(k)}}\sqrt
{\rho_{c_{ij},c_{ij}}^{(k)}}=\sum\limits_{k}t_{k}\sum\limits_{1\leq i\neq
j\leq N}\sqrt{\rho_{c_{0},c_{0}}^{(k)}}\sqrt{\rho_{c_{ij},c_{ij}}^{(k)}}%
\]
Finally, for the third term,%
\[
\sum\limits_{1\leq i\leq N}\sum\limits_{k}t_{k}\rho_{c_{i},c_{i}}^{(k)}%
=\sum\limits_{k}t_{k}\sum\limits_{1\leq i\leq N}\rho_{c_{i},c_{i}}^{(k)}.
\]
Putting everything together, we obtain the bound in the statement:%

\begin{align*}
\mathcal{F}(\rho,\psi)  &  \leq\sum\limits_{k}t_{k}\sum\limits_{1\leq i\neq
j\leq N}|\rho^{(k)}_{c_{i},c_{j}}|-\sum\limits_{k}t_{k}\sum\limits_{1\leq
i\neq j\leq N}\sqrt{\rho_{c_{0},c_{0}}^{(k)}}\sqrt{\rho_{c_{ij},c_{ij}}^{(k)}%
}-(N-2)\sum\limits_{k}t_{k}\sum\limits_{1\leq i\leq N}\rho^{(k)}_{c_{i},c_{i}%
}\\
&  =\sum\limits_{k}t_{k}(\sum\limits_{1\leq i\neq j\leq N}|\rho^{(k)}%
_{c_{i},c_{j}}|-\sum\limits_{1\leq i\neq j\leq N}\sqrt{\rho_{c_{0},c_{0}%
}^{(k)}}\sqrt{\rho_{c_{ij},c_{ij}}^{(k)}}-(N-2)\sum\limits_{1\leq i\leq N}%
\rho^{(k)}_{c_{i},c_{i}})\\
&  \leq2\sqrt{(N-1)}\sum\limits_{k}t_{k}\cdot C_{GME}(\rho^{(k)}%
)=2\sqrt{(N-1)}\cdot C_{GME}(\rho)
\end{align*}

\subsubsection{Bound 2}

As we have already done above, we start with a warm-up case. It will be a
four-qubit state. For a four-qubit state, $|\psi\rangle=|0000\rangle$, we
obtain $|\psi_{1}\rangle=|1000\rangle$, $|\psi_{2}\rangle=|0100\rangle$,
$|\psi_{3}\rangle=|0010\rangle$, and $|\psi_{4}\rangle=|0001\rangle$. We shall
prove that
\[
\mathcal{L}(\rho,\psi_{1})+\mathcal{L}(\rho,\psi_{2})+\mathcal{L}(\rho
,\psi_{3})+\mathcal{L}(\rho,\psi_{4})\leq2\sqrt{2}\cdot C_{GME}(\rho),
\]
where%
\begin{align*}
\mathcal{L}(\rho,\psi_{1})  &  =2(|\rho_{10,11}|+|\rho_{10,13}|+|\rho
_{11,13}|-\sqrt{\rho_{9,9}\rho_{12,12}}-\sqrt{\rho_{9,9}\rho_{14,14}}%
-\sqrt{\rho_{9,9}\rho_{15,15}})-(|\rho_{10,10}|+|\rho_{11,11}|+|\rho
_{13,13}|)\\
&  =L_{11}+L_{12}-L_{13};\\
\mathcal{L}(\rho,\psi_{2})  &  =2(|\rho_{6,7}|+|\rho_{6,13}|+|\rho
_{7,13}|-\sqrt{\rho_{5,5}\rho_{8,8}}-\sqrt{\rho_{5,5}\rho_{14,14}}-\sqrt
{\rho_{5,5}\rho_{15,15}})-(|\rho_{6,6}|+|\rho_{7,7}|+|\rho_{13,13}|)\\
&  =L_{21}+L_{22}-L_{23};\\
\mathcal{L}(\rho,\psi_{3})  &  =2(|\rho_{4,7}|+|\rho_{4,11}|+|\rho
_{7,11}|-\sqrt{\rho_{3,3}\rho_{8,8}}-\sqrt{\rho_{3,3}\rho_{12,12}}-\sqrt
{\rho_{3,3}\rho_{15,15}})-(|\rho_{4,4}|+|\rho_{7,7}|+|\rho_{11,11}|)\\
&  =L_{31}+L_{32}-L_{33};\\
\mathcal{L}(\rho,\psi_{4})  &  =2(|\rho_{4,6}|+|\rho_{4,10}|+|\rho
_{6,10}|-\sqrt{\rho_{2,2}\rho_{8,8}}-\sqrt{\rho_{2,2}\rho_{12,12}}-\sqrt
{\rho_{2,2}\rho_{14,14}})-(|\rho_{4,4}|+|\rho_{6,6}|+|\rho_{10,10}|)\\
&  =L_{41}+L_{42}-L_{43}.
\end{align*}
where $L_{ij}(1\leq i\leq4,1\leq j\leq3)$ depend on the partition. Each
$L_{ij}$ is given in Appendix A. When $\rho$ is a pure state, that is
$\rho=|\phi\rangle\langle\phi|$, for $C_{1}(\phi)$, we have the following:%

\begin{align*}
L_{21}  &  =2(|\phi_{0101}\phi_{1100}|+|\phi_{0110}\phi_{1100}|-|\phi
_{0100}\phi_{1101}|-|\phi_{0100}\phi_{1110}|);\\
L_{22}  &  =(2|\phi_{0101}\phi_{0110}|-2|\phi_{0100}\phi_{0111}|-|\phi
_{0101}|^{2}-|\phi_{0110}|^{2});\\
L_{23}  &  =|\phi_{1100}|^{2};\\
L_{31}  &  =2(|\phi_{0011}\phi_{1010}|+|\phi_{0110}\phi_{1010}|-|\phi
_{0010}\phi_{1011}|-|\phi_{0010}\phi_{1110}|);\\
L_{32}  &  =(2|\phi_{0011}\phi_{0110}|-2|\phi_{0010}\phi_{0111}|-|\phi
_{0011}|^{2}-|\phi_{0110}|^{2});\\
L_{33}  &  =|\phi_{1010}|^{2};\\
L_{41}  &  =2(|\phi_{0011}\phi_{1001}|+|\phi_{0101}\phi_{1001}|-|\phi
_{0001}\phi_{1011}|-|\phi_{0001}\phi_{1101}|);\\
L_{42}  &  =(2|\phi_{0011}\phi_{0101}|-2|\phi_{0001}\phi_{0111}|-|\phi
_{0011}|^{2}-|\phi_{0101}|^{2});\\
L_{43}  &  =|\phi_{1001}|^{2}.
\end{align*}
It is obvious that $L_{22},L_{32},L_{42}\leq0$, $L_{21}+L_{31}+L_{41}%
\leq2\sqrt{2}C_{1}(\rho)$, and $\mathcal{L}(\phi,\psi_{1})-|\phi_{1001}%
|^{2}-|\phi_{1010}|^{2}-|\phi_{1100}|^{2}\leq0$. Hence,%
\[
\mathcal{L}(\phi,\psi_{1})+\mathcal{L}(\phi,\psi_{2})+\mathcal{L}(\phi
,\psi_{3})+\mathcal{L}(\phi,\psi_{4})\leq2\sqrt{2}C_{1}(\phi).
\]
The same holds for $C_{2}(\phi)$, $C_{3}(\phi)$, $C_{4}(\phi)$, $C_{12}(\phi
)$, $C_{13}(\phi)$, and $C_{14}(\phi)$. As a consequence, for a pure state,
\begin{align*}
&  \mathcal{L}(\phi,\psi_{1})+\mathcal{L}(\phi,\psi_{2})+\mathcal{L}(\phi
,\psi_{3})+\mathcal{L}(\phi,\psi_{4})\\
&  \leq2\sqrt{2}\min\{C_{1}(\phi),C_{2}(\phi),C_{3}(\phi),C_{4}(\phi
),C_{12}(\phi),C_{13}(\phi),C_{14}(\phi)\}\\
&  =2\sqrt{2}C_{GME}(\phi).
\end{align*}
If $\rho$ is a mixed state, the convex roof construction is bounded as
\[
2\sqrt{2}C_{GME}(\rho)\geq\inf\limits_{\{p_{i},|\phi_{i}\rangle\}}\sum
_{i}p_{i}(\mathcal{L}(\phi_{i},\psi_{1})+\mathcal{L}(\phi_{i},\psi
_{2})+\mathcal{L}(\phi_{i},\psi_{3})+\mathcal{L}(\phi_{i},\psi_{4})),
\]
where ${\{}p_{i},\phi^{(i)}{\}}$ is any pure state decomposition of $\rho$. Since $C_{GME}(\rho)$ is invariant under local unitaries, for
any choice of a product state $|\psi_{1}\rangle$, $\mathcal{L}(\rho,\psi_{1})$
is a lower bound to $C_{GME}(\rho)$ leading to $2\sqrt{2}C_{GME}(\rho
)\geq(\mathcal{L}(\rho,\psi_{1})+\mathcal{L}(\rho,\psi_{2})+\mathcal{L}%
(\rho,\psi_{3})+\mathcal{L}(\rho,\psi_{4}))$. This concludes the proof for the
four-qubit case.

We are now ready to prove the inequality for a general $N$-qubit state. If%
\[
c_{ijk}:=x_{1}\cdots x_{i-1}x_{i}^{\prime}x_{i+1}\cdots x_{j-1}x_{j}^{\prime
}x_{j+1}\cdots x_{k-1}x_{k}^{\prime}x_{k+1}\cdots x_{N}%
\]
then%
\[
\mathcal{L}(\phi,\psi_{c_{i}})=\sum\limits_{j\neq i,k\neq i,j\neq k}%
2(|\phi_{c_{ij}}\phi_{c_{ik}}|-|\phi_{c_{0}}\phi_{c_{ijk}}|)-(N-3)\sum
\limits_{1\leq j\leq N,j\neq i}|\phi_{c_{ij}}|^{2}.
\]
There are again two cases in close analogy with the previus part.

\noindent\textbf{Case 1. }For any given $\gamma\subset\{1,2,...N\}$ with
$|\{\gamma\}|=1$,%
\begin{align*}
2\sqrt{N-2}C_{\gamma}(\phi)  &  =2\sqrt{N-2}\sqrt{\sum\limits_{j\neq
\gamma,i\neq\gamma,j\neq i}|\phi_{c_{i\gamma}}\phi_{c_{ij}}-\phi_{c_{i}}%
\phi_{c_{ij\gamma}}|^{2}}\\
&  \geq\sum\limits_{j\neq\gamma,i\neq\gamma,j\neq i}|\phi_{c_{i\gamma}}%
\phi_{c_{ij}}-\phi_{c_{i}}\phi_{c_{ij\gamma}}|\geq\sum\limits_{j\neq
\gamma,i\neq\gamma,j\neq i}|\phi_{c_{i\gamma}}\phi_{c_{ij}}|-|\phi_{c_{i}}%
\phi_{c_{ij\gamma}}|,
\end{align*}
Now,%
\begin{align*}
\mathcal{L}(\phi,\psi_{i})  &  =\sum\limits_{j\neq\gamma,i\neq\gamma,j\neq
i}2(|\phi_{c_{ij}}\phi_{c_{i\gamma}}|-|\phi_{c_{i}}\phi_{c_{ij\gamma}}%
|)+\sum\limits_{i\neq\gamma,j\neq\gamma,k\neq\gamma,i\neq j,i\neq k}%
2(|\phi_{c_{ij}}\phi_{c_{ik}}|-|\phi_{c_{i}}\phi_{c_{ijk}}|)\\
&  -(N-3)\sum\limits_{1\leq j\leq N,j\neq i,j\neq\gamma}|\phi_{c_{ij}}%
|^{2}-(N-3)|\phi_{c_{i\gamma}}|^{2}\\
&  =L_{i1}+L_{i2}-L_{i3},
\end{align*}
where%
\begin{align*}
L_{i1}  &  =\sum\limits_{j\neq\gamma,i\neq\gamma,j\neq i}2(|\phi_{c_{ij}}%
\phi_{c_{i\gamma}}|-|\phi_{c_{i}}\phi_{c_{ij\gamma}}|);\\
L_{i2}  &  =\sum\limits_{i\neq\gamma,j\neq\gamma,k\neq\gamma,i\neq j}%
2(|\phi_{c_{ij}}\phi_{c_{ik}}|-|\phi_{c_{i}}\phi_{c_{ijk}}|)-(N-3)\sum
\limits_{1\leq j\leq N,j\neq i}|\phi_{c_{ij}}|^{2}\\
&  \leq\sum\limits_{i\neq\gamma,j\neq\gamma,k\neq\gamma,i\neq j}(|\phi
_{c_{ij}}|^{2}+|\phi_{c_{ik}}|^{2})-(N-3)\sum\limits_{1\leq j\leq N,j\neq
i}|\phi_{c_{ij}}|^{2}\\
&  =(N-3)\sum\limits_{1\leq j\leq N,j\neq i}|\phi_{c_{ij}}|^{2}-(N-3)\sum
\limits_{1\leq j\leq N,j\neq i,j\neq\gamma}|\phi_{c_{ij}}|^{2}\leq0;\\
L_{i3}  &  =(N-3)|\phi_{c_{i\gamma}}|^{2}.
\end{align*}
Thus,%
\begin{align*}
\sum\limits_{1\leq i\leq N,i\neq\gamma}L_{i1}  &  \leq2\sqrt{2}C_{GME}%
(\rho),\\
L_{i2}  &  \leq0,\text{ for }1\leq i\leq N,
\end{align*}
and%
\[
\mathcal{L}(\phi,\psi_{\gamma})-\sum\limits_{1\leq i\leq N,i\neq\gamma}%
L_{i3}\leq0.
\]
So, for pure states,
\begin{align*}
\mathcal{L}(\phi,\psi_{1})+\mathcal{L}(\phi,\psi_{2})+\mathcal{L}(\phi
,\psi_{3})+\mathcal{L}(\phi,\psi_{4})  &  \leq2\sqrt{2}\min\{C_{1}(\phi
),C_{2}(\phi),C_{3}(\phi),C_{4}(\phi),C_{12}(\phi),C_{13}(\phi),C_{14}%
(\phi)\}\\
&  =2\sqrt{2}C_{GME}(\phi).
\end{align*}

\noindent\textbf{Case 2. }For any given $\gamma=\{j_{1},j_{2},...,j_{k}%
\}\subset\{1,2,...N\}$, with $k\geq2$,
\begin{align*}
2\sqrt{N-2}\cdot C_{\gamma}(\phi)  &  \geq2\sqrt{N-2)}\cdot C_{\gamma}(\phi)\\
&  =2\sqrt{N-2}\sqrt{\sum\limits_{l=1}^{k}\sum\limits_{j\neq j_{t}:1\leq t\leq
k}|\phi_{c_{ij_{l}}}\phi_{c_{ij}}-\phi_{c_{i}}\phi_{c_{ij_{l}j}}|^{2}}\geq
\sum\limits_{l=1}^{k}\sum\limits_{j\neq j_{t}:1\leq t\leq k}|\phi_{c_{ij_{l}}%
}\phi_{c_{ij}}|-|\phi_{c_{i}}\phi_{c_{ij_{l}j}}|
\end{align*}
As in the previous case,
\begin{align*}
\mathcal{L}(\phi,\psi_{i})  &  =\sum\limits_{l=1}^{k}\sum\limits_{j\neq
j_{t}:1\leq t\leq k}(|\phi_{c_{ij}}\phi_{c_{ij_{l}}}|-|\phi_{c_{i}}%
\phi_{c_{ijj_{l}}}|)+\sum\limits_{i\neq j,i\neq j_{t},s\neq j_{t},j\neq
j_{t}:1\leq t\leq k}(|\phi_{c_{is}}\phi_{c_{ij}}|-|\phi_{c_{i}}\phi_{c_{ijs}%
}|)\\
&  +\sum\limits_{l\neq t}(|\phi_{c_{ij_{t}}}\phi_{c_{ij_{l}}}|-|\phi_{c_{i}%
}\phi_{c_{ij_{t}j_{l}}}|)-(N-3)\sum\limits_{1\leq j\leq N,j\neq i}%
|\phi_{c_{ij}}|^{2}\\
&  =L_{i1}+L_{i2},
\end{align*}
where $X$ is the summand with $i=j_{l}$ and $j\neq j_{t}$ or $j=j_{l}$ and
$i\neq j_{t}$ ($1\leq t\leq k$); $Y$ is the summand with $i\neq j_{l}$ and
$j\neq j_{l}$ ($l=1,2,...,k$) or $i=j_{l},j=j_{t}$. Then,%
\[
L_{i1}=\sum\limits_{l=1}^{k}\sum\limits_{j\neq j_{t}:1\leq t\leq k}%
(|\phi_{c_{ij}}\phi_{c_{ij_{l}}}|-|\phi_{c_{i}}\phi_{c_{ijj_{l}}}|)
\]
and%
\begin{align*}
L_{i2}  &  =\sum\limits_{i\neq j,i\neq j_{t},s\neq j_{t},j\neq j_{t}:1\leq
t\leq k}(|\phi_{c_{is}}\phi_{c_{ij}}|-|\phi_{c_{i}}\phi_{c_{ijs}}%
|)+\sum\limits_{l\neq t}(|\phi_{c_{ij_{t}}}\phi_{c_{ij_{l}}}|-|\phi_{c_{i}%
}\phi_{c_{ij_{t}j_{l}}}|)\\
&  -(N-3)\sum\limits_{1\leq j\leq N,j\neq i}|\phi_{c_{ij}}|^{2}\\
&  \leq\sum\limits_{i\neq j,i\neq j_{t},s\neq j_{t},j\neq j_{t}:1\leq t\leq
k}\frac{|\phi_{c_{is}}|^{2}+|\phi_{c_{ij}}|^{2}}{2}+\sum\limits_{l\neq t}%
\frac{|\phi_{c_{ij_{l}}}|^{2}+|\phi_{c_{ij_{t}}}|^{2}}{2}-(N-3)\sum
\limits_{1\leq j\leq N,j\neq i}|\phi_{c_{ij}}|^{2}\\
&  \leq(N-k-1)\sum\limits_{i\neq s:1\leq t\leq k}(|\phi_{c_{is}}%
|^{2})+(k-2)\sum\limits_{l}(|\phi_{c_{ij_{l}}}|^{2})\\
&  \leq(N-3)\sum\limits_{i}|\phi_{c_{i}}|^{2}-(N-3)\sum\limits_{i}|\phi
_{c_{i}}|^{2}\\
&  =0.
\end{align*}
For an $n$-qubit state,
\[
\sum\limits_{1\leq i\leq N}\mathcal{L}(\phi,\psi_{i})\leq2\sqrt{N-2}%
\min\limits_{\gamma}\{C_{\gamma}(\phi)\}=2\sqrt{N-2}C_{GME}(\phi).
\]

\subsubsection{Bound 3}

Let $V=\{|\chi_{1}\rangle,...,|\chi_{4}\rangle\}$ be a set of product states
in $\mathcal{H}$,where $|\chi_{1}\rangle=|0011\rangle$, $|\chi_{2}%
\rangle=|0101\rangle$, $|\chi_{3}\rangle=|0110\rangle$, and $|\chi_{4}%
\rangle=|1010\rangle$. Let $K_{1}=\{|\chi_{2}\rangle,|\chi_{3}\rangle
,|\chi_{4}\rangle\}$, $K_{2}=\{|\chi_{1}\rangle,|\chi_{3}\rangle\}$,
$K_{3}=\{|\chi_{1}\rangle,|\chi_{2}\rangle,|\chi_{4}\rangle\}$, and
$K_{4}=\{|\chi_{1}\rangle,|\chi_{3}\rangle\}$. Here, $s=3$. Details of the
next steps are in Appendix B. If $\rho$ is a pure state, $\rho=|\phi
\rangle\langle\phi|$, we can prove that Bound 3 is $\mathcal{T}(\rho,\chi
)\leq3\sqrt{2}\cdot C_{GME}(\rho)$. For any given $\gamma\subset\{1,2,...N\}$
such that $|\{\gamma\}|=1$, take, \emph{e.g.}, $\gamma=\{1\}$, \emph{i.e.},
the case $C_{1}(\rho)$. Let $\mathcal{T}(\rho,\chi)=X+Y$. The sum giving $X$
involves terms $|\chi_{\alpha}\rangle$ and $|\chi_{\beta}\rangle$ that are
different in qubit 1; on the other hand, the terms of $Y$ are the states
$|\chi_{\alpha}\rangle$ and $|\chi_{\beta}\rangle$ that are different in the
other qubits, except qubit 1. This is $X\leq3\sqrt{2}C_{1}(\rho)$ and $Y\leq0$
(see Appendix B.1). So $\mathcal{T}(\rho,\chi)=X+Y\leq3\sqrt{2}C_{1}(\rho)$.
We can use the same process to get $\mathcal{T}(\rho,\chi)=X+Y\leq3\sqrt
{2}C_{2}(\rho)$, $\mathcal{T}(\rho,\chi)=X+Y\leq3\sqrt{2}C_{3}(\rho)$, and
$\mathcal{T}(\rho,\chi)=X+Y\leq3\sqrt{2}C_{4}(\rho)$.

Now if $|\{\gamma\}|=2$, take $\gamma=\{1,2\}$ as the example, $\emph{i.e.}$,
$C_{12}(\rho)$. Again, let $\mathcal{T}(\rho,\chi)=X+Y$.The sum giving $X$
involves terms $|\chi_{\alpha}\rangle$ and $|\chi_{\beta}\rangle$ that are
different in only one of the $\gamma$ qubits; the terms of $Y$ are defined
analogously to the previous case. Hence, $X\leq3\sqrt{2}C_{12}(\rho)$ and
$Y\leq0$ (see Appendix B.2). We obtain $\mathcal{T}(\rho,\chi)=X+Y\leq
3\sqrt{2}C_{12}(\rho)$, and, by the same process, $\mathcal{T}(\rho
,\chi)=X+Y\leq3\sqrt{2}C_{13}(\rho)$, $\mathcal{T}(\rho,\chi)=X+Y\leq3\sqrt
{2}C_{14}(\rho)$, and $\mathcal{T}(\rho,\chi)\leq3\sqrt{2}C_{GME}(\rho)$.

For a mixed state $\rho$, the bound is given by using the convex roof
construction. For the general case, the sum giving $\mathcal{T}(\rho,\chi)$
simplifies because only two qubits of $|\chi_{\alpha}\rangle$ and
$|\chi_{\beta}\rangle$ are different. For any given $\gamma\subset
\{1,2,...N\}$, $\mathcal{T}(\rho,\chi)$ can be subdivided into two addends,
which we will denote by $X$ and $Y$. If $|\{\gamma\}|=1$, the sum giving $X$
involves terms $|\chi_{\alpha}\rangle$ and $|\chi_{\beta}\rangle$ that are
different in the $\gamma$ qubit; on the other hand, the terms of $Y$ are the
states $|\chi_{\alpha}\rangle$ and $\chi_{\beta}\rangle$ that are different in
the other qubits except for the $\gamma$ qubit; the case $|\{\gamma\}|>1$ is
an easy generalization. From this,%

\[
X=\sum\limits_{|\chi_{\alpha\rangle\in V}}\sum\limits_{A}(|\langle\chi
_{\alpha}|\rho|\chi_{\beta}\rangle|-\sqrt{\langle\chi_{\alpha}|\otimes
\langle\chi_{\beta}|\Pi_{\alpha\beta}\rho^{\otimes2}\Pi_{\alpha\beta}%
|\chi_{\alpha}\rangle\otimes|\chi_{\beta}\rangle})
\]
and%
\[
Y=\sum\limits_{|\chi_{\alpha\rangle\in V}}\sum\limits_{B}(|\langle\chi
_{\alpha}|\rho|\chi_{\beta}\rangle|-\sqrt{\langle\chi_{\alpha}|\otimes
\langle\chi_{\beta}|\Pi_{\alpha\beta}\rho^{\otimes2}\Pi_{\alpha\beta}%
|\chi_{\alpha}\rangle\otimes|\chi_{\beta}\rangle})-(s-s_{0})\sum
\limits_{a}\langle\chi_{\alpha}|\rho|\chi_{\alpha}\rangle.
\]
When $\rho$ is a pure state, $\rho=|\phi\rangle\langle\phi|$, if
$|\chi_{\alpha}\rangle$ and $|\chi_{\beta}\rangle$ are different in the
$\gamma$ qubit only, the value $|\langle\chi_{\alpha}|\rho|\chi_{\beta}%
\rangle-\sqrt{\langle\chi_{\alpha}|\otimes\langle\chi_{\beta}|\Pi_{\alpha
\beta}\rho^{\otimes2}\Pi_{\alpha\beta}|\chi_{\alpha}\rangle\otimes|\chi
_{\beta}\rangle}|$ is one of the terms of $C_{\gamma}(\rho)$ and the maximum
number of terms in $X$ is less than $2s^{2}$; when all states in $K_{\alpha}$
belong to $A$ for any $\chi_{\alpha}\in V$, and for any $\alpha$, $|K_{\alpha
}|=s$, the maximum number of terms in $X$ is $2(s+s-1+\cdots+1)=s(s+1)\leq
2s^{2}$. So, $X\leq\sqrt{2}sC_{\gamma}(\rho)$. Concerning $Y$, for every
$|\chi_{\alpha}\rangle$ in $V$, the maximum number of terms in the sum is
$s-s_{0}$. For any $\alpha$, the maximum number of the terms $|\langle
\chi_{\alpha}|\rho|\chi_{\beta}\rangle|$ is $s$, but the minimum number of
terms $|\langle\chi_{\alpha}|\rho|\chi_{\beta}\rangle|$ in $X$ is $s_{0}$, and
so the maximum number of terms in $Y$ is $s-s_{0}$. Hence, $Y\leq0$ and
$X+Y\leq\sqrt{2}sC_{\gamma}(\rho)$. For a mixed state $\rho$, the convex roof
construction is bounded as
\[
\sqrt{2}sC_{GME}(\rho)\geq\inf\limits_{\{p_{i},|\phi_{i}\rangle\}}\sum
_{i}p_{i}\mathcal{T}(\phi_{i},\psi),
\]
where ${\{}p_{i},\phi^{(i)}{\}}$ is any pure state decomposition of $\rho$.
Consequently we can get the desired result that $\mathcal{T}(\rho,\chi
)\leq\sqrt{2}sC_{GME}(\rho)$.

\section{Examples}

In this section, we illustrate our main result with some explicit examples.
The first one is useful to clarify the bounds:

\begin{example}
\emph{Given }$|\phi\rangle=|0000\rangle$\emph{, we obtain }$|\phi_{1}%
\rangle=|1000\rangle$\emph{, }$|\phi_{2}\rangle=|0100\rangle$\emph{, }%
$|\phi_{3}\rangle=|0010\rangle$\emph{, and }$|\phi_{4}\rangle=|0001\rangle
$\emph{, by applying the bit flip operation. The bound in Eq. (\ref{eqth1})
is} $\mathcal{L}(\rho,\phi_{1})+\mathcal{L}(\rho,\phi_{2})+\mathcal{L}%
(\rho,\phi_{3})+\mathcal{L}(\rho,\phi_{4})\leq2\sqrt{2}\cdot C_{GME}(\rho)$,
\emph{where}%
\begin{align*}
\mathcal{L}(\rho,\phi_{1})  &  =2(\left\vert \rho_{10,11}\right\vert
+\left\vert \rho_{10,13}\right\vert +\left\vert \rho_{11,13}\right\vert
-\sqrt{\rho_{9,9}\rho_{12,12}}-\sqrt{\rho_{9,9}\rho_{14,14}}-\sqrt{\rho
_{9,9}\rho_{15,15}})-(\rho_{10,10}+\rho_{11,11}+\rho_{13,13});\\
\mathcal{L}(\rho,\phi_{2})  &  =2(\left\vert \rho_{6,7}\right\vert +\left\vert
\rho_{6,13}\right\vert +\left\vert \rho_{7,13}\right\vert -\sqrt{\rho
_{5,5}\rho_{8,8}}-\sqrt{\rho_{5,5}\rho_{14,14}}-\sqrt{\rho_{5,5}\rho_{15,15}%
})-(\rho_{6,6}+\rho_{7,7}+\rho_{13,13});\\
\mathcal{L}(\rho,\phi_{3})  &  =2(\left\vert \rho_{4,7}\right\vert +\left\vert
\rho_{4,11}\right\vert +\left\vert \rho_{7,11}\right\vert -\sqrt{\rho
_{3,3}\rho_{8,8}}-\sqrt{\rho_{3,3}\rho_{12,12}}-\sqrt{\rho_{3,3}\rho_{15,15}%
})-(\rho_{4,4}+\rho_{7,7}+\rho_{11,11});\\
\mathcal{L}(\rho,\phi_{4})  &  =2(\left\vert \rho_{4,6}\right\vert +\left\vert
\rho_{4,10}\right\vert +\left\vert \rho_{6,10}\right\vert -\sqrt{\rho
_{2,2}\rho_{8,8}}-\sqrt{\rho_{2,2}\rho_{12,12}}-\sqrt{\rho_{2,2}\rho_{14,14}%
})-(\rho_{4,4}+\rho_{6,6}+\rho_{10,10}).
\end{align*}
\emph{For the bound in Eq. (\ref{eqth2}), let us fix }$|v_{1}\rangle
=|0011\rangle$\emph{, }$|v_{2}\rangle=|0101\rangle$\emph{, }$|v_{3}%
\rangle=|0110\rangle$\emph{, and }$|v_{4}\rangle=|1010\rangle$\emph{. Let
}$K_{1}=\{|v_{2}\rangle,|v_{3}\rangle,|v_{4}\rangle\}$\emph{, }$K_{2}%
=\{|v_{1}\rangle,|v_{3}\rangle\}$\emph{, }$K_{3}=\{|v_{1}\rangle,|v_{2}%
\rangle,|v_{4}\rangle\}$\emph{, and }$K_{4}=\{|v_{1}\rangle,|v_{3}\rangle
\}$\emph{ (}$s=3$\emph{). Then }$2(\left\vert \rho_{4,6}\right\vert
+\left\vert \rho_{4,7}\right\vert +\left\vert \rho_{4,11}\right\vert
+\left\vert \rho_{6,7}\right\vert +\left\vert \rho_{7,11}\right\vert
-\sqrt{\rho_{2,2}\rho_{8,8}}-\sqrt{\rho_{8,8}\rho_{5,5}}-\sqrt{\rho_{3,3}%
\rho_{5,5}}-\sqrt{\rho_{3,3}\rho_{12,12}}-\sqrt{\rho_{3,3}\rho_{15,15}%
})-2(\rho_{4,4}+\rho_{6,6}+\rho_{7,7}+\rho_{11,11})\leq\sqrt{6}\cdot
C_{GME}(\rho)$.\emph{ }
\end{example}

\begin{example}
\emph{Let us consider a two-parameter four-qubit state given by a mixture of
the identity matrix, the }$W$\emph{ state, and the anti-}$W$\emph{ state: }%
\begin{equation}
\rho=\frac{1-a-b}{32}I_{32}+a|\widetilde{W}\rangle\langle\widetilde
{W}|+b|W\rangle\langle W| \label{fourqubits}%
\end{equation}
\emph{with }$|W\rangle=\frac{1}{\sqrt{5}}(|00001\rangle+|00010\rangle
+|00100\rangle+|01000\rangle+|10000\rangle)$\emph{ and }$|\widetilde{W}%
\rangle=\frac{1}{\sqrt{5}}(|11110\rangle+|11101\rangle+|11011\rangle
+|10111\rangle+|01111\rangle)$. \emph{Fig. (\ref{fig1a}) illustrates the GME
area detected by the Bound 1 (Section \ref{res}) and Eq. (III) in
\cite{Huber10}, respectively. The area detected by the former is visibly
larger.}
%TCIMACRO{\FRAME{fhFU}{2.3671in}{1.5995in}{0pt}{\Qcb{The entanglement area of
%the density matrix $\rho$ in Eq. \ref{fourqubits} detected by Bound 1. The
%area is above the lowest line. The area detected by Eq. (III) in
%\cite{Huber10} is above the middle line. }}{\Qlb{fig1a}}{figes.eps}%
%width 2.3671in;  height 1.5995in;  depth 0pt;  original-width 3.4395in;
%original-height 2.3103in;  cropleft "0";  croptop "1";  cropright "1";
%cropbottom "0";  filename 'figes.eps';file-properties "XNPEU";}} }%

\emph{Our Bound 1 is as follows:}%
\begin{align*}
4\sqrt
{2}C_{GME}(\rho)\geq &  2(|\rho_{2,3}|+|\rho_{2,5}|+|\rho_{2,9}|+|\rho_{2,17}|+|\rho_{3,5}%
|+|\rho_{3,9}|+|\rho_{3,17}|+|\rho_{5,9}|+|\rho_{5,17}|+|\rho_{9,17}|\\
&  -\sqrt{\rho_{1,1}\rho_{4,4}}-\sqrt{\rho_{1,1}\rho_{6,6}}-\sqrt{\rho
_{1,1}\rho_{10,10}}-\sqrt{\rho_{1,1}\rho_{18,18}}-\sqrt{\rho_{1,1}\rho_{7,7}%
}\\
&  -\sqrt{\rho_{1,1}\rho_{11,11}}-\sqrt{\rho_{1,1}\rho_{19,19}}-\sqrt
{\rho_{1,1}\rho_{13,13}}-\sqrt{\rho_{1,1}\rho_{21,21}}-\sqrt{\rho_{1,1}%
\rho_{25,25}})\\
&  -3(\rho_{2,2}+\rho_{3,3}+\rho_{5,5}+\rho_{9,9}+\rho_{17,17});
\end{align*}%
\begin{align*}
4\sqrt
{2}C_{GME}(\rho)\geq &  2(|\rho_{16,24}|+|\rho_{16,28}|+|\rho_{16,30}|+|\rho_{16,31}|+|\rho
_{24,28}|+|\rho_{24,30}|+|\rho_{24,31}|+|\rho_{28,30}|+|\rho_{28,31}%
|+|\rho_{30,31}|\\
&  -\sqrt{\rho_{32,32}\rho_{8,8}}-\sqrt{\rho_{32,32}\rho_{12,12}}-\sqrt
{\rho_{32,32}\rho_{14,14}}-\sqrt{\rho_{32,32}\rho_{15,15}}-\sqrt{\rho
_{32,32}\rho_{20,20}}\\
&  -\sqrt{\rho_{32,32}\rho_{22,22}}-\sqrt{\rho_{32,32}\rho_{23,23}}-\sqrt
{\rho_{32,32}\rho_{26,26}}-\sqrt{\rho_{32,32}\rho_{27,27}}-\sqrt{\rho
_{32,32}\rho_{29,29}})\\
&  -3(\rho_{16,16}+\rho_{24,24}+\rho_{28,28}+\rho_{30,30}+\rho_{31,31}%
).
\end{align*}

\emph{The above two equations give bounds to }$C_{GME}(\rho)$. \emph{From
these, we get }$\frac{67b+35a-35}{64}\leq4\sqrt{2}C_{GME}(\rho)$\emph{ and
}$\frac{67a+35b-35}{64}\leq4\sqrt{2}C_{GME}(\rho)$\emph{, respectively. \ The
entanglement area above the lowest line is obtained by taking }$\frac
{67b+35a-35}{64}>0$\emph{ or }$\frac{67a+35b-35}{64}>0$\emph{. The
entanglement area is the union set of} $\{(a,b)|\frac{67b+35a-35}{64}>0\}$
\emph{and }$\{(a,b)|\frac{67a+35b-35}{64}>0\}$ \emph{. The bound of Eq. (III)
in \cite{Huber10} gives instead}%
\begin{align*}
4\sqrt{2}C_{GME}(\rho)\geq &  2(|\rho_{2,3}|+|\rho_{2,5}|+|\rho_{2,9}|+|\rho_{2,17}|+|\rho_{3,5}%
|+|\rho_{3,9}|+|\rho_{3,17}|+|\rho_{5,9}|+|\rho_{5,17}|+|\rho_{9,17}|)\\
&  -3(2\sqrt{\rho_{1,1}\rho_{4,4}}+2\sqrt{\rho_{1,1}\rho_{6,6}}+2\sqrt
{\rho_{1,1}\rho_{10,10}}+2\sqrt{\rho_{1,1}\rho_{18,18}}\\
&  +2\sqrt{\rho_{1,1}\rho_{7,7}}+2\sqrt{\rho_{1,1}\rho_{11,11}}+2\sqrt
{\rho_{1,1}\rho_{19,19}}+2\sqrt{\rho_{1,1}\rho_{13,13}}\\
&  +2\sqrt{\rho_{1,1}\rho_{21,21}}+2\sqrt{\rho_{1,1}\rho_{25,25}}+\rho
_{2,2}+\rho_{3,3}+\rho_{5,5}+\rho_{9,9}+\rho_{17,17});
\end{align*}%
\begin{align*}
4\sqrt
{2}C_{GME}(\rho)\geq &  2(|\rho_{16,24}|+|\rho_{16,28}|+|\rho_{16,30}|+|\rho_{16,31}|+|\rho
_{24,28}|+|\rho_{24,30}|+|\rho_{24,31}|+|\rho_{28,30}|+|\rho_{28,31}%
|+|\rho_{30,31}|)\\
&  -3(2\sqrt{\rho_{32,32}\rho_{8,8}}+2\sqrt{\rho_{32,32}\rho_{12,12}}%
+2\sqrt{\rho_{32,32}\rho_{14,14}}+2\sqrt{\rho_{32,32}\rho_{15,15}}\\
&  +2\sqrt{\rho_{32,32}\rho_{20,20}}+2\sqrt{\rho_{32,32}\rho_{22,22}}%
+2\sqrt{\rho_{32,32}\rho_{23,23}}+2\sqrt{\rho_{32,32}\rho_{26,26}}\\
&  +2\sqrt{\rho_{32,32}\rho_{27,27}}+2\sqrt{\rho_{32,32}\rho_{29,29}}%
+\rho_{16,16}+\rho_{24,24}+\rho_{28,28}+\rho_{30,30}+\rho_{31,31}).
\end{align*}
\end{example}

\end{widetext}

%BeginExpansion
\begin{figure}
[h]
\begin{center}
\includegraphics[
height=1.5995in,
width=2.3671in
]%
{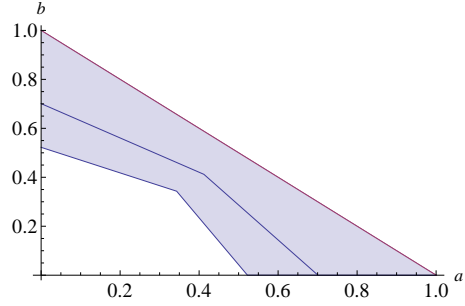}%
\caption{The entanglement area of the density matrix $\rho$ in Eq.
\ref{fourqubits} detected by Bound 1. The area is above the lowest line. The
area detected by Eq. (III) in \cite{Huber10} is above the middle line. }%
\label{fig1a}%
\end{center}
\end{figure}
%EndExpansion

\emph{From these equations, we have }$\frac{75a+107b-75}{32}$\emph{ and
}$\frac{75b+107a-75}{32}$\emph{, respectively. The entanglement area above the
middle line is obtained by }$\frac{75a+107b-75}{32}>0$\emph{ or }%
$\frac{75b+107b-75}{32}>0$\emph{. }

\begin{example}
\emph{Let us consider the one-parameter four-qubit state }$\rho=\frac{1-a}%
{16}I_{16}+a|\phi\rangle\langle\phi|$\emph{, with }$|\phi\rangle=\frac{1}%
{2}(|0011\rangle+|0101\rangle+|0110\rangle+|1010\rangle)$\emph{. Bound 3, Eq.
(\ref{eqth2})) gives the bound }$2(\left\vert \rho_{4,6}\right\vert
+\left\vert \rho_{4,11}\right\vert -\sqrt{\rho_{2,2}\rho_{8,8}}-\sqrt
{\rho_{3,3}\rho_{12,12}})-(\rho_{4,4}+\rho_{6,6}+\rho_{11,11})$. \emph{This
shows that GME area is }$a>\frac{7}{11}$\emph{. On the other hand, by making
use of a criterion in \cite{Huber11a} or our Eq. (\ref{eqth1}), we can find
that there is GME for }$a>\frac{9}{11}$\emph{.}
\end{example}

In the following examples, the method in \cite{Gao10} and Eq. (III) in
\cite{Huber10} can not detect entanglement at all.

\begin{example}
\emph{Let us consider the one-parameter three-qutrit state }$\rho=\frac
{1-a}{27}I_{27}+a|\phi\rangle\langle\phi|$\emph{, with }$|\phi\rangle=\frac
{1}{\sqrt{3}}(|012\rangle+|021\rangle+|111\rangle)$\emph{. Bound 1, Eq.
(\ref{result}) indicates that there is GME for any }$a>\frac{1}{4}$\emph{.
Also, to give a four-qubit example, let }$\rho=\frac{1-a}{16}I_{16}%
+a|\phi\rangle\langle\phi|$\emph{, with }$|\phi\rangle=\frac{1}{2}%
(|0000\rangle+|1100\rangle+|1001\rangle+|1010\rangle)$\emph{. Our result
detects GME for }$a>\frac{5}{9}$\emph{. }
\end{example}

\begin{example}
\emph{Let us consider the one-parameter four-qubit state} $\rho=\frac{1-a}%
{16}I_{16}+a|\phi\rangle\langle\phi|$\emph{, with }$|\phi\rangle=\frac
{1}{\sqrt{5}}(|0000\rangle+|1100\rangle+|1001\rangle+|1010\rangle
+|0110\rangle)$. \emph{By applying Eq. (\ref{result}), we can see that the GME
area is }$a>\frac{25}{41}=0.60976$\emph{. However, Eq. (\ref{eqth2}) (Bound 3)
detects GME for }$a>\frac{45}{61}=0.7377$\emph{. }
\end{example}

\begin{example}
\emph{Let us consider the one-parameter four-qubits state }$\rho=\frac
{1-a}{16}I_{16}+a|\phi_{0112}^{4}\rangle\langle\phi_{0112}^{4}|$\emph{,
}$|\phi_{0112}^{4}\rangle=\frac{1}{\sqrt{6}}(|1100\rangle+|0110\rangle
+|1001\rangle+|0101\rangle+|1010\rangle+|0011\rangle)$. \emph{The density
matrix }$\rho$\emph{ is a mixture of white noise and the Dicke state
\cite{Dicke}. By applying Bound 2 given in Eq. (\ref{eqth1}), }$a>\frac{9}%
{17}$\emph{.}
\end{example}

\section{Conclusions}

We have studied genuine multipartite entanglement of quantum states. We have
given an alternative definition of pure state GME-concurrence based on the
expectation value of an observable with respect to a two-fold copy of the
state under consideration. This definition has the advantage of being physical
accessible (\emph{e.g.}, with the use of twin photons \cite{Guhne,Walborn}). We have
then proposed several analytical lower bounds for GME-concurrence. Such bounds
are also given as expectation values of some observable. On the basis of the
bounds, we have obtained entanglement criteria that can be used to detect GME
for states of generic dimension. We have reported examples in which the
criteria perform better than the previously known methods.

\noindent\emph{Note added}. Recently, we became aware of Ref.\cite{marc}, where
the authors also derive similar results.

\bigskip

{\noindent\textbf{Acknowledgment.}} ZHM is supported by NSF of China
(10901103) and by the Foundation of China Scholarship Council (2010831012).
JLC is supported by NSF of China (10975075) and the Fundamental Research Funds
for the Central Universities. SS is supported by the Royal Society. Part of
this work has been carried out while SS was visiting the Institute of Natural
Sciences at Shanghai Jiao-Tong University. The financial support of this
institution is gratefully acknowledged. While finishing our manuscript we
became aware of \cite{marc}, where the authors also derive similar results. We
are indebted to Marcus Huber for carefully reading earlier drafts and for many
important comments that helped improving the paper.

\begin{widetext}

\appendix{}

\section{Proof of Bound 2:\ details}%

\[%
\begin{tabular}
[c]{ll}%
$1|234$ \\
$L_{21}=2(|\rho_{6,13}|+|\rho_{7,13}|-\sqrt{\rho_{5,5}\rho_{14,14}}-\sqrt
{\rho_{5,5}\rho_{15,15}})$,
$L_{22}=(2|\rho_{6,7}|-2\sqrt{\rho_{5,5}\rho_{8,8}}-|\rho_{6,6}|-|\rho
_{7,7}|)$, & \\
$L_{23}=|\rho_{13,13}|$ ,
$L_{31}=2(|\rho_{4,11}|+|\rho_{7,11}|-\sqrt{\rho_{3,3}\rho_{12,12}}-\sqrt
{\rho_{3,3}\rho_{15,15}})$, & \\
$L_{32}=(2|\rho_{4,7}|-2\sqrt{\rho_{3,3}\rho_{8,8}}-|\rho_{4,4}|-|\rho
_{7,7}|)$ ,
$L_{33}=|\rho_{11,11}|$, & \\
$L_{41}=2(|\rho_{4,10}|+|\rho_{6,10}|-\sqrt{\rho_{2,2}\rho_{12,12}}-\sqrt
{\rho_{2,2}\rho_{14,14}})$, & \\
$L_{42}=(2|\rho_{4,6}|-2\sqrt{\rho_{2,2}\rho_{8,8}}-|\rho_{4,4}|-|\rho
_{6,6}|)$ ,
$L_{43}=|\rho_{10,10}|$. & %
\end{tabular}
\ \
\]%

\[%
\begin{tabular}
[c]{ll}%
$2|134$ \\
$L_{11}=2(|\rho_{10,11}|+|\rho_{11,13}|-\sqrt{\rho_{9,9}\rho_{12,12}}-\sqrt{\rho_{9,9}\rho_{15,15}})$, &\\
$L_{12}=(2|\rho_{10,13}|-2\sqrt{\rho_{9,9}\rho_{14,14}}%
-|\rho_{10,10}|-|\rho_{13,13}|)$ ,
 $L_{13}=|\rho_{11,11}|$,&\\
 $L_{31}=2(|\rho_{4,7}|+|\rho_{7,11}|-\sqrt
{\rho_{3,3}\rho_{8,8}}-\sqrt{\rho_{3,3}\rho_{15,15}})$,&\\
$L_{32}=(2|\rho_{4,11}|-2\sqrt{\rho_{3,3}\rho_{12,12}}-|\rho
_{4,4}|-|\rho_{11,11}|)$,
$L_{33}=|\rho_{7,7}|$,&\\
$L_{41}=2(|\rho_{4,6}|+|\rho_{6,10}|-\sqrt
{\rho_{2,2}\rho_{8,8}}-\sqrt{\rho_{2,2}\rho_{14,14}})$,&\\
$L_{42}=(2|\rho_{4,10}|-2\sqrt{\rho_{2,2}\rho_{12,12}}-|\rho
_{4,4}|-|\rho_{10,10}|)$,
$L_{43}=|\rho_{6,6}|.$ & %
\end{tabular}
\ \
\]%

\[%
\begin{tabular}
[c]{ll}%
$3|124$ \\
$L_{11}=2(|\rho_{10,13}|+|\rho_{11,13}|-\sqrt{\rho_{9,9}\rho_{14,14}}%
-\sqrt{\rho_{9,9}\rho_{15,15}}),$\\
$L_{12}=(2|\rho_{10,11}|-2\sqrt{\rho_{9,9}\rho_{12,12}}-|\rho_{10,10}%
|-|\rho_{11,11}|)$ ,
$L_{13}=|\rho_{13,13}|$ ,\\
$L_{21}=2(|\rho_{6,7}|+|\rho_{7,13}|-\sqrt{\rho_{5,5}\rho_{8,8}}-\sqrt
{\rho_{5,5}\rho_{15,15}})$,\\
$L_{22}=(2|\rho_{6,13}|-2\sqrt{\rho_{5,5}\rho_{14,14}}-|\rho_{6,6}%
|-|\rho_{13,13}|)$ ,
$L_{23}=|\rho_{7,7}|$, &\\
$L_{41}=2(|\rho_{4,6}|+|\rho_{4,10}|-\sqrt{\rho_{2,2}\rho_{8,8}}-\sqrt
{\rho_{2,2}\rho_{12,12}})$, &\\
$L_{42}=(2|\rho_{6,10}|-2\sqrt{\rho_{2,2}\rho_{14,14}}-|\rho_{6,6}%
|-|\rho_{10,10}|)$,
$L_{43}=|\rho_{4,4}|$. %
\end{tabular}
\ \
\]

\[%
\begin{tabular}
[c]{ll}%
 $4|123$\\
 $L_{11}=2(|\rho_{10,13}|+|\rho
_{11,13}|-\sqrt{\rho_{9,9}\rho_{14,14}}-\sqrt{\rho_{9,9}\rho_{15,15}})$,& \\
 $L_{12}=(2|\rho_{10,11}|-2\sqrt{\rho_{9,9}\rho_{12,12}%
}-|\rho_{10,10}|-|\rho_{11,11}|)$, $L_{13}=|\rho_{13,13}|$, &\\
 $L_{21}=2(|\rho_{6,7}|+|\rho_{7,13}|-\sqrt
{\rho_{5,5}\rho_{8,8}}-\sqrt{\rho_{5,5}\rho_{15,15}})$,&\\
 $L_{22}=(2|\rho_{6,13}|-2\sqrt{\rho_{5,5}\rho_{14,14}%
}-|\rho_{6,6}|-|\rho_{13,13}|)$,$L_{23}=|\rho_{7,7}|$,&\\
 $L_{31}=2(|\rho_{4,7}|+|\rho_{4,11}|-\sqrt
{\rho_{3,3}\rho_{8,8}}-\sqrt{\rho_{3,3}\rho_{12,12}})$, &\\
 $L_{32}=(2|\rho_{7,11}|-2\sqrt{\rho_{3,3}\rho_{15,15}%
}-|\rho_{7,7}|-|\rho_{11,11}|)$, $L_{33}=|\rho_{4,4}|.$%
\end{tabular}
\ \
\]

\[%
\begin{tabular}
[c]{ll}%
$12|34$ \\
$L_{11}=2(|\rho_{10,13}|+|\rho_{11,13}|-\sqrt{\rho_{9,9}\rho_{14,14}}%
-\sqrt{\rho_{9,9}\rho_{15,15}})$, \\
$L_{12}=(2|\rho_{10,11}|-2\sqrt{\rho_{9,9}\rho_{12,12}}-|\rho_{10,10}%
|-|\rho_{11,11}|)$ ,$L_{13}=|\rho_{13,13}|$, &\\
$L_{21}=2(|\rho_{6,7}|+|\rho_{7,13}|-\sqrt{\rho_{5,5}\rho_{8,8}}-\sqrt
{\rho_{5,5}\rho_{15,15}})$, & \\
$L_{22}=(2|\rho_{6,13}|-2\sqrt{\rho_{5,5}\rho_{14,14}}-|\rho_{6,6}%
|-|\rho_{13,13}|)$ ,$L_{23}=|\rho_{7,7}|$, &\\
$L_{31}=2(|\rho_{4,7}|+|\rho_{7,11}|-\sqrt{\rho_{3,3}\rho_{8,8}}-\sqrt
{\rho_{3,3}\rho_{15,15}})$, & \\
$L_{32}=(2|\rho_{4,11}|-2\sqrt{\rho_{3,3}\rho_{12,12}}-|\rho_{4,4}%
|-|\rho_{11,11}|)$ ,$L_{33}=|\rho_{7,7}|$, & \\
$L_{41}=2(|\rho_{4,6}|+|\rho_{6,10}|-\sqrt{\rho_{2,2}\rho_{8,8}}-\sqrt
{\rho_{2,2}\rho_{12,12}})$, & \\
$L_{42}=(2|\rho_{4,10}|-2\sqrt{\rho_{2,2}\rho_{14,14}}-|\rho_{4,4}%
|-|\rho_{10,10}|)$ ,$L_{43}=|\rho_{6,6}|$. & %
\end{tabular}
\ \
\]

\[%
\begin{tabular}
[c]{ll}%
 $13|24$\\
  $L_{11}=2(|\rho_{10,11}|+|\rho
_{11,13}|-\sqrt{\rho_{9,9}\rho_{12,12}}-\sqrt{\rho_{9,9}\rho_{15,15}})$,\\
 $L_{12}=(2|\rho_{10,13}|-2\sqrt{\rho_{9,9}\rho_{14,14}%
}-|\rho_{10,10}|-|\rho_{13,13}|)$,$L_{13}=|\rho_{11,11}|$,\\
 $L_{21}=2(|\rho_{6,7}|+|\rho_{6,13}|-\sqrt
{\rho_{5,5}\rho_{8,8}}-\sqrt{\rho_{5,5}\rho_{14,14}})$,\\
 $L_{22}=(2|\rho_{7,13}|-2\sqrt{\rho_{5,5}\rho_{15,15}%
}-|\rho_{7,7}|-|\rho_{13,13}|)$,$L_{23}=|\rho_{6,6}|$,\\
 $L_{31}=2(|\rho_{4,11}|+|\rho_{7,11}|-\sqrt
{\rho_{3,3}\rho_{12,12}}-\sqrt{\rho_{3,3}\rho_{15,15}})$,\\
 $L_{32}=(2|\rho_{4,7}|-2\sqrt{\rho_{3,3}\rho_{8,8}}%
-|\rho_{4,4}|-|\rho_{7,7}|)$,$L_{33}=|\rho_{11,11}|$,\\
 $L_{41}=2(|\rho_{4,6}|+|\rho_{6,10}|-\sqrt
{\rho_{2,2}\rho_{8,8}}-\sqrt{\rho_{2,2}\rho_{12,12}})$,\\
 $L_{42}=(2|\rho_{4,10}|-2\sqrt{\rho_{2,2}\rho_{14,14}%
}-|\rho_{4,4}|-|\rho_{10,10}|)$,$L_{43}=|\rho_{6,6}|.$%
\end{tabular}
\ \
\]

\[%
\begin{tabular}
[c]{l}%
$14|23$\\
$L_{11}=2(|\rho_{10,11}|+|\rho_{10,13}|-\sqrt{\rho_{9,9}\rho_{12,12}}%
-\sqrt{\rho_{9,9}\rho_{14,14}})$,\\
$L_{12}=(2|\rho_{11,13}|-2\sqrt{\rho_{9,9}\rho_{15,15}}-|\rho_{11,11}%
|-|\rho_{13,13}|)$,$L_{13}=|\rho_{10,10}|$,\\
$L_{21}=2(|\rho_{6,7}|+|\rho_{7,13}|-\sqrt{\rho_{5,5}\rho_{8,8}}-\sqrt
{\rho_{5,5}\rho_{15,15}})$,\\
$L_{22}=(2|\rho_{6,13}|-2\sqrt{\rho_{5,5}\rho_{14,14}}-|\rho_{6,6}%
|-|\rho_{13,13}|)$,$L_{23}=|\rho_{7,7}|$,\\
$L_{31}=2(|\rho_{4,7}|+|\rho_{7,11}|-\sqrt{\rho_{3,3}\rho_{8,8}}-\sqrt
{\rho_{3,3}\rho_{15,15}})$,\\
$L_{32}=(2|\rho_{4,11}|-2\sqrt{\rho_{3,3}\rho_{12,12}}-|\rho_{4,4}%
|-|\rho_{11,11}|)$,$L_{33}=|\rho_{7,7}|$,\\
$L_{41}=2(|\rho_{4,10}|+|\rho_{6,10}|-\sqrt{\rho_{2,2}\rho_{14,14}}-\sqrt
{\rho_{2,2}\rho_{12,12}})$,\\
$L_{42}=(2|\rho_{4,6}|-2\sqrt{\rho_{2,2}\rho_{8,8}}-|\rho_{4,4}|-|\rho
_{6,6}|)$,$L_{43}=|\rho_{10,10}|.$%
\end{tabular}
\
\]

\section{Proof of Bound 3:\ details}

\subsection{$C_{1}(\rho)$}%

\begin{align*}
\mathcal{T}(\rho,\chi)  &  =|\langle\chi_{1}|\rho|\chi_{2}\rangle
|-\sqrt{\langle\chi_{1}|\otimes\langle\chi_{2}|\Pi_{12}\rho^{\otimes2}\Pi
_{12}|\chi_{1}\rangle\otimes|\chi_{2}\rangle}+|\langle\chi_{1}|\rho|\chi
_{3}\rangle|-\sqrt{\langle\chi_{1}|\otimes\langle\chi_{3}|\Pi_{13}%
\rho^{\otimes2}\Pi_{13}|\chi_{1}\rangle\otimes|\chi_{3}\rangle}\\
&  +|\langle\chi_{1}|\rho|\chi_{4}\rangle|-\sqrt{\langle\chi_{1}%
|\otimes\langle\chi_{4}|\Pi_{14}\rho^{\otimes2}\Pi_{14}|\chi_{1}\rangle
\otimes|\chi_{4}\rangle}+|\langle\chi_{2}|\rho|\chi_{1}\rangle|-\sqrt
{\langle\chi_{2}|\otimes\langle\chi_{1}|\Pi_{12}\rho^{\otimes2}\Pi_{12}%
|\chi_{2}\rangle\otimes|\chi_{1}\rangle}\\
&  +|\langle\chi_{2}|\rho|\chi_{3}\rangle|-\sqrt{\langle\chi_{2}%
|\otimes\langle\chi_{3}|\Pi_{23}\rho^{\otimes2}\Pi_{23}|\chi_{2}\rangle
\otimes|\chi_{3}\rangle}+|\langle\chi_{3}|\rho|\chi_{1}\rangle|-\sqrt
{\langle\chi_{3}|\otimes\langle\chi_{1}|\Pi_{13}\rho^{\otimes2}\Pi_{13}%
|\chi_{3}\rangle\otimes|\chi_{1}\rangle}\\
&  +|\langle\chi_{3}|\rho|\chi_{2}\rangle|-\sqrt{\langle\chi_{3}%
|\otimes\langle\chi_{2}|\Pi_{23}\rho^{\otimes2}\Pi_{23}|\chi_{3}\rangle
\otimes|\chi_{2}\rangle}+|\langle\chi_{3}|\rho|\chi_{4}\rangle|-\sqrt
{\langle\chi_{3}|\otimes\langle\chi_{4}|\Pi_{34}\rho^{\otimes2}\Pi_{34}%
|\chi_{3}\rangle\otimes|\chi_{4}\rangle}\\
&  +|\langle\chi_{4}|\rho|\chi_{1}\rangle|-\sqrt{\langle\chi_{4}%
|\otimes\langle\chi_{1}|\Pi_{14}\rho^{\otimes2}\Pi_{14}|\chi_{4}\rangle
\otimes|\chi_{1}\rangle}+|\langle\chi_{4}|\rho|\chi_{3}\rangle|-\sqrt
{\langle\chi_{4}|\otimes\langle\chi_{3}|\Pi_{34}\rho^{\otimes2}\Pi_{34}%
|\chi_{4}\rangle\otimes|\chi_{3}\rangle})\\
&  =2(|\langle\chi_{1}|\rho|\chi_{2}\rangle|-\sqrt{\langle\chi_{1}%
|\otimes\langle\chi_{2}|\Pi_{12}\rho^{\otimes2}\Pi_{12}|\chi_{1}\rangle
\otimes|\chi_{2}\rangle}+|\langle\chi_{1}|\rho|\chi_{3}\rangle|-\sqrt
{\langle\chi_{1}|\otimes\langle\chi_{3}|\Pi_{13}\rho^{\otimes2}\Pi_{13}%
|\chi_{1}\rangle\otimes|\chi_{3}\rangle}\\
&  +|\langle\chi_{1}|\rho|\chi_{4}\rangle|-\sqrt{\langle\chi_{1}%
|\otimes\langle\chi_{4}|\Pi_{14}\rho^{\otimes2}\Pi_{14}|\chi_{1}\rangle
\otimes|\chi_{4}\rangle}+|\langle\chi_{2}|\rho|\chi_{3}\rangle|-\sqrt
{\langle\chi_{2}|\otimes\langle\chi_{3}|\Pi_{23}\rho^{\otimes2}\Pi_{23}%
|\chi_{2}\rangle\otimes|\chi_{3}\rangle}\\
&  +|\langle\chi_{3}|\rho|\chi_{4}\rangle|-\sqrt{\langle\chi_{3}%
|\otimes\langle\chi_{4}|\Pi_{34}\rho^{\otimes2}\Pi_{34}|\chi_{3}\rangle
\otimes|\chi_{4}\rangle}\\
&  =2(|\phi_{0011}\phi_{0101}|-|\phi_{0001}\phi_{0111}|+|\phi_{0011}%
\phi_{0110}|-|\phi_{0010}\phi_{1110}|+|\phi_{0011}\phi_{1010}|-|\phi
_{0010}\phi_{1011}|+|\phi_{0110}\phi_{0101}|\\
&  -|\phi_{0100}\phi_{0111}|+|\phi_{0110}\phi_{1010}|-|\phi_{0010}\phi
_{1110}|)-2(|\phi_{0011}|^{2}+|\phi_{0101}|^{2}+|\phi_{0110}|^{2}+|\phi
_{1010}|^{2})\\
&  \leq3\sqrt{2}\cdot C_{GME}(\rho);
\end{align*}%
\begin{align*}
X  &  =|\langle\chi_{1}|\rho|\chi_{4}\rangle|-\sqrt{\langle\chi_{1}%
|\otimes\langle\chi_{4}|\Pi_{14}\rho^{\otimes2}\Pi_{14}|\chi_{1}\rangle
\otimes|\chi_{4}\rangle}+|\langle\chi_{3}|\rho|\chi_{4}\rangle|-\sqrt
{\langle\chi_{3}|\otimes\langle\chi_{4}|\Pi_{34}\rho^{\otimes2}\Pi_{34}%
|\chi_{3}\rangle\otimes|\chi_{4}\rangle}\\
&  +|\langle\chi_{4}|\rho|\chi_{1}\rangle|-\sqrt{\langle\chi_{4}%
|\otimes\langle\chi_{1}|\Pi_{14}\rho^{\otimes2}\Pi_{14}|\chi_{4}\rangle
\otimes|\chi_{1}\rangle}+|\langle\chi_{4}|\rho|\chi_{3}\rangle|-\sqrt
{\langle\chi_{4}|\otimes\langle\chi_{3}|\Pi_{34}\rho^{\otimes2}\Pi_{34}%
|\chi_{4}\rangle\otimes|\chi_{3}\rangle}\\
&  =2(|\phi_{0011}\phi_{1010}|-|\phi_{1011}\phi_{0010}|+|\phi_{0110}%
\phi_{1010}|-|\phi_{1110}\phi_{0010}|)\\
&  \leq2(|\phi_{0011}\phi_{1010}-\phi_{1011}\phi_{0010}|+|\phi_{0110}%
\phi_{1010}-\phi_{1110}\phi_{0010}|)\\
&  \leq3\sqrt{2}C_{1}(\rho);
\end{align*}%
\begin{align*}
Y  &  =|\langle\chi_{1}|\rho|\chi_{2}\rangle|-\sqrt{\langle\chi_{1}%
|\otimes\langle\chi_{2}|\Pi_{12}\rho^{\otimes2}\Pi_{12}|\chi_{1}\rangle
\otimes|\chi_{2}\rangle}+|\langle\chi_{1}|\rho|\chi_{3}\rangle|-\sqrt
{\langle\chi_{1}|\otimes\langle\chi_{3}|\Pi_{13}\rho^{\otimes2}\Pi_{13}%
|\chi_{1}\rangle\otimes|\chi_{3}\rangle}\\
&  +|\langle\chi_{2}|\rho|\chi_{1}\rangle|-\sqrt{\langle\chi_{2}%
|\otimes\langle\chi_{1}|\Pi_{12}\rho^{\otimes2}\Pi_{12}|\chi_{2}\rangle
\otimes|\chi_{1}\rangle}+|\langle\chi_{2}|\rho|\chi_{3}\rangle|-\sqrt
{\langle\chi_{2}|\otimes\langle\chi_{3}|\Pi_{23}\rho^{\otimes2}\Pi_{23}%
|\chi_{2}\rangle\otimes|\chi_{3}\rangle}\\
&  +|\langle\chi_{3}|\rho|\chi_{1}\rangle|-\sqrt{\langle\chi_{3}%
|\otimes\langle\chi_{1}|\Pi_{13}\rho^{\otimes2}\Pi_{13}|\chi_{3}\rangle
\otimes|\chi_{1}\rangle}+|\langle\chi_{3}|\rho|\chi_{2}\rangle|-\sqrt
{\langle\chi_{3}|\otimes\langle\chi_{2}|\Pi_{23}\rho^{\otimes2}\Pi_{23}%
|\chi_{3}\rangle\otimes|\chi_{2}\rangle}\\
&  =2(|\phi_{0011}\phi_{0101}|-|\phi_{0001}\phi_{0111}|+|\phi_{0011}%
\phi_{0110}|-|\phi_{0010}\phi_{0111}|+|\phi_{0110}\phi_{0101}|-|\phi
_{0100}\phi_{0111}|)\\
&  -2(|\phi_{0011}|^{2}+|\phi_{0101}|^{2}+|\phi_{0110}|^{2}+|\phi_{1010}%
|^{2})\\
&  \leq2(|\phi_{0011}\phi_{0101}|+|\phi_{0011}\phi_{0110}|+|\phi_{0110}%
\phi_{0101}|)-2(|\phi_{0011}|^{2}+|\phi_{0101}|^{2}+|\phi_{0110}|^{2}%
+|\phi_{1010}|^{2})\\
&  \leq0.
\end{align*}

\subsection{$C_{12}(\rho)$}%

\begin{align*}
X  &  =|\langle\chi_{1}|\rho|\chi_{2}\rangle|-\sqrt{\langle\chi_{1}%
|\otimes\langle\chi_{2}|\Pi_{12}\rho^{\otimes2}\Pi_{12}|\chi_{1}\rangle
\otimes|\chi_{2}\rangle}+|\langle\chi_{1}|\rho|\chi_{3}\rangle|-\sqrt
{\langle\chi_{1}|\otimes\langle\chi_{3}|\Pi_{13}\rho^{\otimes2}\Pi_{13}%
|\chi_{1}\rangle\otimes|\chi_{3}\rangle}\\
&  +|\langle\chi_{1}|\rho|\chi_{4}\rangle|-\sqrt{\langle\chi_{1}%
|\otimes\langle\chi_{4}|\Pi_{14}\rho^{\otimes2}\Pi_{14}|\chi_{1}\rangle
\otimes|\chi_{4}\rangle}+|\langle\chi_{2}|\rho|\chi_{1}\rangle|-\sqrt
{\langle\chi_{2}|\otimes\langle\chi_{1}|\Pi_{12}\rho^{\otimes2}\Pi_{12}%
|\chi_{2}\rangle\otimes|\chi_{1}\rangle}\\
&  +|\langle\chi_{3}|\rho|\chi_{1}\rangle|-\sqrt{\langle\chi_{3}%
|\otimes\langle\chi_{1}|\Pi_{13}\rho^{\otimes2}\Pi_{13}|\chi_{3}\rangle
\otimes|\chi_{1}\rangle}+|\langle\chi_{4}|\rho|\chi_{1}\rangle|-\sqrt
{\langle\chi_{4}|\otimes\langle\chi_{1}|\Pi_{14}\rho^{\otimes2}\Pi_{14}%
|\chi_{4}\rangle\otimes|\chi_{1}\rangle}\\
&  =2(|\phi_{0011}\phi_{0101}|-|\phi_{0001}\phi_{0111}|+|\phi_{0011}%
\phi_{0110}|-|\phi_{0010}\phi_{0111}|+|\phi_{0011}\phi_{1010}|-|\phi
_{1011}\phi_{0010}|)\\
&  \leq2(|\phi_{0011}\phi_{0101}-\phi_{0001}\phi_{0111}|+|\phi_{0011}%
\phi_{0110}-\phi_{0010}\phi_{0111}|+|\phi_{0011}\phi_{1010}-\phi_{1011}%
\phi_{0010}|)\\
&  \leq3\sqrt{2}C_{12}(\rho);
\end{align*}%
\begin{align*}
Y  &  =|\langle\chi_{2}|\rho|\chi_{3}\rangle|-\sqrt{\langle\chi_{2}%
|\otimes\langle\chi_{3}|\Pi_{23}\rho^{\otimes2}\Pi_{23}|\chi_{2}\rangle
\otimes|\chi_{3}\rangle}+|\langle\chi_{3}|\rho|\chi_{2}\rangle|-\sqrt
{\langle\chi_{3}|\otimes\langle\chi_{2}|\Pi_{23}\rho^{\otimes2}\Pi_{23}%
|\chi_{3}\rangle\otimes|\chi_{2}\rangle}\\
&  +|\langle\chi_{3}|\rho|\chi_{4}\rangle|-\sqrt{\langle\chi_{3}%
|\otimes\langle\chi_{4}|\Pi_{34}\rho^{\otimes2}\Pi_{34}|\chi_{3}\rangle
\otimes|\chi_{4}\rangle}+|\langle\chi_{4}|\rho|\chi_{3}\rangle|-\sqrt
{\langle\chi_{4}|\otimes\langle\chi_{3}|\Pi_{34}\rho^{\otimes2}\Pi_{34}%
|\chi_{4}\rangle\otimes|\chi_{3}\rangle}\\
&  =2(|\phi_{0110}\phi_{0101}|-|\phi_{0100}\phi_{0111}|+|\phi_{0110}%
\phi_{1010}|-|\phi_{1110}\phi_{0010}|)\\
&  -2(|\phi_{0011}|^{2}+|\phi_{0101}|^{2}+|\phi_{0110}|^{2}+|\phi_{1010}%
|^{2})\\
&  \leq2(|\phi_{0110}\phi_{0101}|+|\phi_{0110}\phi_{1010}|)-2(|\phi
_{0011}|^{2}+|\phi_{0101}|^{2}+|\phi_{0110}|^{2}+|\phi_{1010}|^{2})\\
&  \leq0.
\end{align*}

\end{widetext}


\begin{thebibliography}{99}                                                                                               %


\bibitem {Raussendorf}R. Raussendorf and H.-J. Briegel, \emph{Phys. Rev.
Lett.} \textbf{86}, 5188 (2001).

\bibitem {Schauer}S. Schauer, M. Huber, and B. C. Hiesmayr, \emph{Phys. Rev.
A} \textbf{82}, 062311 (2010).

\bibitem {Markham}D. Markham and B. C. Sanders, \emph{Phys. Rev. A}
\textbf{78}, 042309 (2008).

\bibitem {Dur00}W. D\"{u}r, G. Vidal, and J. I. Cirac, \emph{Phys. Rev. A}
\textbf{62}, 062314 (2000).

\bibitem {Barreiro10}J. T. Barreiro, P. Schindler, O. G\"{u}hne, T. Monz, M.
Chwalla, C. F. Roos, M. Hennrich, R. Blatt, \emph{Nat. Phys.} 6,
\textbf{943} (2010).

\bibitem {Guhe11}B. Jungnitsch, T. Moroder, and O. G\"{u}hne, \emph{Phys. Rev.
Lett.} \textbf{106}, 190502 (2011).

\bibitem {Bell}J.-D. Bancal, N. Gisin, Y.-C. Liang, and S. Pironio,
\emph{Phys. Rev. Lett.} \textbf{106}, 250404 (2011).

\bibitem {Horodecki09}R. Horodecki, P. Horodecki, M. Horodecki, and K.
Horodecki, \emph{Rev. Mod. Phys.} \textbf{81}, 865 (2009).

\bibitem {Guhne09}O. G\"{u}hne, G. Toth, \emph{Phys. Rep.} \textbf{474}, 1(2009).

\bibitem {Coffman00}V. Coffman, J. Kundu and W. K. Wootters, \emph{Phys. Rev.
A} \textbf{61}, 052306 (2000).

\bibitem {Ma11}Z.-H. Ma, Z.-H. Chen, J.-L. Chen, C. Spengler, A. Gabriel, and
M. Huber, \emph{Phys. Rev. A} \textbf{83}, 062325 (2011).

\bibitem {Aolita06}L. Aolita, F. Mintert, \emph{Phys. Rev. Lett.} \textbf{97}, 050501(2006).

\bibitem {Huber10}M. Huber, F. Mintert, A. Gabriel, and B. C. Hiesmayr,
\emph{Phys. Rev. Lett.} \textbf{104}, 210501 (2010).

\bibitem {Guhne}O. G\"{u}hne and M. P. Seevinck, \emph{New J. Phys.}
\textbf{12}, 053002 (2010).

\bibitem {Huber11a}M. Huber, P. Erker, H. Schimpf, A. Gabriel, and B. C.
Hiesmayr, \emph{Phys. Rev. A} \textbf{83}, 040301(R) (2011).

\bibitem {Gao10}T. Gao, Y. Hong, \emph{Phys. Rev. A} \textbf{82}, 062113 (2010).

\bibitem {Dicke}R. H. Dicke, \emph{Phys. Rev.} \textbf{93}, 99 (1954).

\bibitem {Walborn}S. P. Walborn, P. H. Souto Ribeiro, L. Davidovich, F.
Mintert, and A. Buchleitner, \emph{Nature(London)} \textbf{440}, 1022 (2006).

\bibitem {marc}J. Wu, H. Kampermann, D. Bru\ss , C. Kl\"{o}ckl, M. Huber,
Experimentally feasible lower bounds on measures of multipartite entanglement,
\emph{arXiv:1205.3119}.
\end{thebibliography}
\end{document}